\documentclass[twocolumn]{aastex631}

\usepackage{amsmath}

 \newcommand{ \Rearth }{R_{\oplus}}

\submitjournal{ApJ}

\shorttitle{Monosilane worlds}

\begin{document}

\title{
Monosilane Worlds: Sub-Neptunes with Atmospheres Shaped by Reduced Magma Oceans
}

\footnote{\today}

\correspondingauthor{Yuichi Ito}
\email{yuichi.ito.kkyr@gmail.com}

\author[0000-0002-0598-3021]{Yuichi Ito}
\affiliation{Division of Science, National Astronomical Observatory of Japan, 2-21-1 Osawa, Mitaka, Tokyo 181-8588, Japan}
\affiliation{Department of Physics and Astronomy, University College London 
Gower Street, WC1E 6BT London, United Kingdom}

\author[0000-0001-8477-2523]{Tadahiro Kimura}
\affiliation{UTokyo Organization for Planetary Space Science (UTOPS), University of Tokyo, Hongo, Bunkyo-ku, Tokyo 113-0033, Japan}
\affiliation{Kapteyn Astronomical Institute, University of Groningen Landleven 12, 9747 AD, Groningen, Netherlands}

\author[0000-0003-3290-6758]{Kazumasa Ohno}
\affiliation{Division of Science, National Astronomical Observatory of Japan, 2-21-1 Osawa, Mitaka, Tokyo 181-8588, Japan}

\author[0000-0002-2786-0786]{Yuka Fujii}\affiliation{Division of Science, National Astronomical Observatory of Japan, 2-21-1 Osawa, Mitaka, Tokyo 181-8588, Japan}
\affiliation{Graduate Institute for Advanced Studies, SOKENDAI, 2-21-1 Osawa, Mitaka, Tokyo 181-8588, Japan}

\author[0000-0002-5658-5971]{Masahiro Ikoma}
\affiliation{Division of Science, National Astronomical Observatory of Japan, 2-21-1 Osawa, Mitaka, Tokyo 181-8588, Japan}
\affiliation{Graduate Institute for Advanced Studies, SOKENDAI, 2-21-1 Osawa, Mitaka, Tokyo 181-8588, Japan}
\affiliation{Department of Earth and Planetary Science, University of Tokyo, Hongo, Bunkyo-ku, Tokyo 113-0033, Japan}

\begin{abstract}
High-precision infrared spectroscopic measurements now enable detailed characterization of sub-Neptune atmospheres, potentially providing constraints on their interiors. Motivated by this, atmospheric models have been developed to explore chemical interactions between hydrogen-dominated atmospheres and possibly underlying magma oceans with various redox states. 
Recent models have predicted monosilane (SiH$_4$) as a potential atmospheric species derived from magma oceans in sub-Neptunes, but suggested that it is highly depleted in 
the observable atmospheric layers. Here, we propose that SiH$_4$ can persist throughout the atmospheres of sub-Neptunes with FeO-free reduced magma oceans
by considering the dissolution of H$_2$O into the magma oceans, a factor not accounted for in previous models. We construct a one-dimensional atmospheric model to simulate the chemical equilibrium composition of hydrogen-dominated atmospheres overlying FeO-free magma oceans, incorporating H-O-Si chemistry.
Our results show that the dissolution of H$_2$O enhances the  SiH$_4$ molar fraction to levels of 0.1--10~\%, preventing it from reverting to silicates in the upper atmospheric layers. We find that SiH$_4$-rich atmospheres can exist across a broad parameter space with ground temperatures of 2000--6000~K and hydrogen pressures of 10$^2$--10$^5$~bar.
We discuss that SiH$_4$-rich atmospheres could contain the other silanes but lack C-/N-/O-bearing species. 
The detection of SiH$_4$ in future observations of sub-Neptunes would provide compelling evidence for the presence of a rocky core with a reduced magma ocean. 
However, the accuracy of our model is limited by the lack of data on the non-ideal behavior and radiative properties of SiH$_4$, highlighting the need for further numerical and laboratory investigations.
\end{abstract}

\section{Introduction}
 \label{sec:intro}
Sub-Neptunes are a class of exoplanets with radii smaller than Neptune and densities lower than rocky planets. They are typically categorized into two primary types: one with a thick hydrogen-dominated atmosphere and a rocky core, and another with a water-rich interior, reflecting different formation histories \citep[e.g.,][]{Jin+2018,Izidoro2022, Burn+2024}. Although mass and radius measurements alone often lead to degeneracies in determining their bulk composition, atmospheric characterization is expected to provide additional constraints to resolve this degeneracy \citep[e.g.,][]{Valencia+07,Adams+08,Miller-Ricci+2009}. Today, the advent of the James Webb Space Telescope (JWST), with its high precision and broad wavelength coverage in atmospheric spectroscopy, has improved our ability to constrain the interior composition of sub-Neptunes \citep[e.g.,][]{Madhusudhan+2023, Shorttle+2024}.

The thick hydrogen-dominated atmospheres of sub-Neptunes can exert a strong blanketing effect making their possible rocky surfaces hot enough to melt and vaporize. The global molten silicate layers, known as magma oceans, can store volatiles and alter the atmospheric composition \citep[e.g.,][]{Chachan+2018, Kite+2020}. 
Similar atmosphere-magma interactions have been extensively studied for rocky planets in the Solar System.
In particular, the partitioning of volatile elements such as H, C, N, O and S between gases and silicate melts is known to be strongly influenced by
the redox state of magma oceans 
 \citep[e.g.,][]{French1966,Frost1979,Holloway1981,Holloway+1994,Hirschmann+2012,ARMSTRONG+2015,Gaillard+2022}.
  For example, previous studies on proto-Earth have proposed that an oxidized magma ocean could lead to the generation of abundant H$_2$O via reactions with nebula-origin hydrogen \citep{Sasaki1990}, while 
 a reduced magma ocean containing metallic Fe could form an H$_2$-dominated atmosphere through its interaction with solid-derived H$_2$O \citep{Kuramoto+1996}.
 The redox state itself is thought to be primarily controlled by the oxidation state of Fe (i.e., metallic Fe, FeO and Fe$_2$O$_3$) and their fractions in a magma ocean \citep[e.g.,][and references therein]{Frost+2008}. 
Given that the oxygen fugacity (a measure of how reduced or oxidized rock is) of meteorites and terrestrial basalts in the Solar System spans over 10 orders of magnitude \citep{Righter+2016}, the redox state of exoplanets' magma oceans, including those of sub-Neptunes, may also exhibit substantial diversity. 

For sub-Neptunes, current theoretical models have incorporated chemical interactions between magma oceans and hydrogen-rich atmospheres, including the vaporization of magma oceans and the dissolution of volatile species into them. These models have demonstrated how  magma oceans influence the atmospheric composition under different redox conditions \citep[e.g.,][]{Kite+2020, Schlichting+2022, Charnoz+2023, Shorttle+2024, Tian+2024, Seo+2024}. 
For example, they have shown that an
oxidized magma ocean with an Earth-like redox state tends to increase the atmospheric H$_2$O/H$_2$ ratio \citep{Kite+2020} and CO$_2$/CO ratio \citep{Tian+2024}, while decreasing the atmospheric C/O ratio \citep{Seo+2024}. In contrast, a reduced magma ocean can lead to the depletion of atmospheric NH$_3$ due to the high solubility of nitrogen into reduced magma oceans \citep{Shorttle+2024}.
The influence of carbon and sulfur content in  magma oceans and the role of innermost iron core have also been investigated \citep{Tian+2024,Schlichting+2022}. Besides volatile species, refractory-element-bearing gas species such as metal hydrides can also contribute to the sub-Neptunes' atmospheric composition via reactions between rocky vapors and hydrogen \citep{Charnoz+2023,Misener+2023}.

Monosilane, SiH$_4$, has recently been predicted as a possible gas species in sub-Neptune atmospheres by chemical equilibrium calculations \citep[][]{Charnoz+2023,Misener+2023}.
Since SiH$_4$ originates from the vaporization of SiO$_2$ in molten silicates, its detection in atmospheric observations of sub-Neptunes would provide compelling evidence for the presence of rocky cores.
 Not only the equilibrium calculations but laboratory experiments using a laser-heated diamond-anvil cell have also indicated the formation of SiH$_4$ and H$_2$O in SiO$_2$-H$_2$ system at 2$\times$10$^4$~bar and 1700~K \citep{Shinozaki+2014}, and in MgSiO$_3$-H$_2$ system at 3.6$\times$10$^4$~bar and 2000~K \citep{Shinozaki+2016}. 
Similar experiments have also indicated the decomposition of Mg$_2$SiO$_4$ and
  SiO$_2$ dissolving into H$_2$ in Mg$_2$SiO$_4$-H$_2$ system at 2.5--15$\times$10$^4$~bar and 1400--1500~K \citep{Shinozaki+2013}.   
Additionally, SiH$_4$ has already  been predicted 
to exist in the deep regions of Jupiter's and Saturn's atmospheres by chemical equilibrium calculations
\citep[e.g.,][]{Fegley+1994, Visscher+2010}.

For SiH$_4$ to be produced in abundant quantities within sub-Neptune atmospheres, reduced magma oceans are likely required, as
SiH$_4$ is highly reactive with oxygen \citep{HARTMAN+1987} and experimental studies indicating the formation of SiH$_4$ have been reported only in FeO-free silicates-H$_2$ systems \citep{Shinozaki+2014,Shinozaki+2016}.
This is also consistent with previous work by \citet{Charnoz+2023} who showed SiH$_4$-rich atmospheres. Although their assumption of an Earth-like FeO-containing magma ocean appears to contradict the requirement for reduced magma, this discrepancy arises because
their model did not account for oxygen buffered by the coexistence of metallic Fe and FeO in the magma ocean, 
a factor considered in other oxidized magma ocean models \citep[e.g.,][]{Kite+2020,Seo+2024}. Instead, they assumed that all metals and oxygen in the atmosphere originated from the vaporization of molten rocky oxides, imposing an elemental conservation relation between oxygen and vaporized metals in their chemical equilibrium calculations. This assumption leads to the depletion of oxygen at high hydrogen partial pressures, as shown in their results \citep[see Fig.B1--4 in][]{Charnoz+2023}.
Consequently, their model showed that 
SiH$_4$ becomes the third most abundant species after H$_2$ and H$_2$O under oxygen-depleted conditions with hydrogen partial pressure above 10$^4$~bar. Thus, their results suggest that atmospheres equilibrating with SiH$_4$-rich conditions would correspond to those with reduced magma oceans.  

Additionally, \citet{Misener+2023} demonstrated a SiH$_4$-rich atmosphere overlying an FeO-free (i.e., reduced) pure SiO$_2$ magma ocean through their chemical equilibrium calculations, assuming the same elemental conservation relation as that used in \citet{Charnoz+2023}. 
They showed that SiH$_4$ was abundant near the base of the atmosphere but was depleted in its upper layers. This finding is consistent with \citet{Falco+2024}, who simulated the  atmospheric vertical structure
 based on the equilibrium calculation of \citet{Charnoz+2023}. Despite SiH$_4$'s high volatility \citep[boiling temperature at 1~atm is 161~K,][]{CRC2014}, its depletion occurs because SiH$_4$ is chemically  favorable in reduced environments with high hydrogen pressure, which prevents it from reverting to refractory Si-oxides. 
This depletion makes its detection challenging through atmospheric spectroscopy \citep{Falco+2024}.  

However, the dissolution of H$_2$O into magma oceans, which has been overlooked in the previous studies demonstrating SiH$_4$-rich atmospheres, could lead to further depleting oxygen and enhancing reducing conditions.
Experimental studies have consistently demonstrated that water is highly soluble in silicate melts \citep[e.g., Table 3 in][and reference therein]{Papale1997}, and the dissolution of H$_2$O into
FeO-free silicate melts
 has also been observed \citep[e.g.,][]{Kennedy1962,Holtz+2000,NOVELLA+2017}. For sub-Neptunes, the dissolution of H$_2$O into magma oceans has been considered in  current magma-atmosphere interaction models for H-O-C-N chemistry \citep[e.g.,][]{Kite+2020,Shorttle+2024,Seo+2024} but not for the H-O-Si system.
This process acts as a reservoir of oxygen, even in the form of H$_2$O, which may create conditions that allow SiH$_4$ to remain abundant throughout sub-Neptune atmospheres.

In an attempt to address the hypothesis, this study explores the possibility of SiH$_4$-rich atmospheres on sub-Neptunes, considering the dissolution of H$_2$O into magma oceans. 
The remainder of this paper is organized as follows. In Section \ref{sec:mod}, we describe our atmospheric model and numerical setup. In Section \ref{sec:res}, we present SiH$_4$ abundance as a function of  surface temperature and pressure under reduced conditions that may have been present on sub-Neptunes. We discuss the planet property to have atmospheric SiH$_4$ and the caveats of our model in Section \ref{sec:disc}.
Finally, we summarize our results in Section \ref{sec:sum}.
 
\section{Model} \label{sec:mod}
\begin{figure}[t]
 \begin{minipage}{0.5\textwidth}
    \begin{center}
\includegraphics[width=\textwidth]{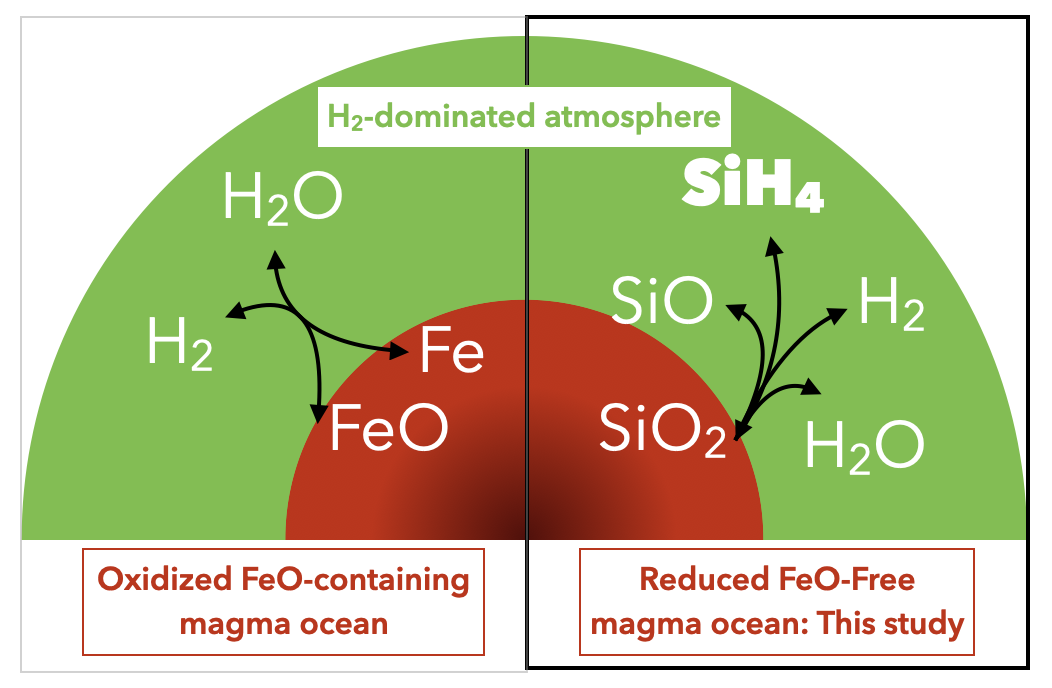}
   \end{center}
 \end{minipage}
\caption{
Schematic illustration of chemical  interaction between an hydrogen-dominated atmosphere and a magma ocean in sub-Neptune with an oxidized FeO-containing magma ocean (left panel) and a reduced FeO-free magma ocean. Note that the focus of this study is the chemical interaction in sub-Neptune with a reduced FeO-free magma ocean. See the text for  details.
}
\label{fig:atom_sch}
\end{figure}
In this study, we model a hydrogen-dominated atmosphere overlying an 
 FeO-free magma ocean on a sub-Neptune (see Fig.~\ref{fig:atom_sch}), representing a highly reduced condition. Such a condition might arise if planets  formed with reduced materials, experienced efficient differentiation of their innermost iron cores, and had iron cores isolated from the outer silicates, 
 as discussed later in Section~\ref{sec:d_r}. 
 In our model, we adopt a chemical network similar to that of a previous study on Fe{O}-free magma oceans \citep[][]{Misener+2023}, with the additional consideration of the dissolution of H$_2$O into magma oceans.
 The atmospheric chemistry is governed by SiO$_2$, which differs from models considering the oxygen buffer of Fe-oxides in oxidized magma oceans \citep[e.g.,][]{Kite+2020}, as shown in Fig.~\ref{fig:atom_sch}.

Regarding the core properties of a sub-Neptune,
we assume a core mass of $4M_\oplus$ and an iron mass fraction equivalent to that of Earth's core as a representative case. This choice is based on \citet{Rogers+2021} who analyzed exoplanets with radii $\leq 4R_\oplus$ and orbital periods $\leq 100$ days. Their findings show that the core mass distribution peaks at $\sim 4~M_\oplus$, and that the iron core fraction is consistent with that of an Earth-like core. Additionally, we assume a core radius of 1.4~$R_\oplus$ which  is derived from the mass-radius relationship for a 4~$M_\oplus$ rocky core with an Earth-like iron fraction of 33 wt\% \citep{Fortney+2007}.

We assume that the atmosphere is in hydrostatic 
and chemical equilibrium, and the core is covered with a global magma ocean due to the strong atmospheric blanketing effect.
{For atmospheres containing abundant SiH$_4$, modeling an equilibrium temperature profile is currently challenging due to the lack of available opacity data for SiH$_4$ at wavelengths shorter than 2 microns, as further discussed in Sec.~\ref{sec:d_t}. 
Given this limitation, we adopt a simplified atmospheric temperature structure  to determine the vertical profiles of  composition}
from the ground to the optical {photospheric} radius, which is set
at 0.1~bar \citep[e.g.,][]{Fortney+2019}. 
The temperature profile is assumed to be adiabatic 
from the ground to the tropopause and isothermal 
above the tropopause{, for simplicity}. Therefore, the isothermal layer is assumed to have the same temperature as the tropopause temperature, $T_\mathrm{rcb}$.
We treat the ground pressure of hydrogen, $P_\mathrm{H_2,gr}$, ground temperature, $T_\mathrm{gr}$, and tropopause pressure, $P_\mathrm{rcb}$, as parameters, exploring ranges of 100 bar to 0.1 Mbar, 2000~K to 6000~K, and 1 bar to 100 bar, respectively.
{
The parameter range explored in our study are also listed in Table~\ref{tab:param}.
This simplified temperature profile, together with its parameterization, allows us to isolate and clarify the dependence of atmospheric composition on these thermal conditions.}
Using the equation of state of an H-He mixture with a He mass fraction, $Y_\mathrm{He}$,
of 0.275 from \citet{Chabrier+2021} and the equation of state of ideal gas for the other gas species with the additive volume law, we determine the temperature-pressure-density profile of the atmosphere.
The H$_2$ molar fraction of the H-He mixture is approximately given as, $\chi_\mathrm{H_2}=(2-2Y_\mathrm{He})/(2-Y_\mathrm{He})\sim 0.84$. 
We discuss later the influence of  simplifications regarding the equation of state and 
 temperature profiles in Sec.\ref{sec:d_cm}.

{
\begin{table}[b]
	\centering
	\caption{Explored parameter range}
	\label{tab:example_table}
 	\begin{tabular}{cc} 
	Parameters & Range of parameter values \\
    \hline
     $P_\mathrm{H_2, gr}$ & 10$^2$--10$^5$~bar \\
     $T_\mathrm{gr}$ & 2000--6000~K \\
     $P_\mathrm{rcb}$ & 1--100~bar \\
	\end{tabular}
\label{tab:param}
\end{table}
}

\subsection{Atmospheric chemistry}

\subsubsection{Gas phase reactions}
For the atmospheric chemistry at the ground, which represents the magma-atmosphere interface, we consider the vaporization of SiO$_2$ on the magma ocean, followed by thermal dissociation, namely,
\begin{equation}
\mathrm{SiO}_{2} (\mathrm{l}) \rightleftharpoons \mathrm{SiO} + \frac{1}{2} \mathrm{O_2}, \tag{R1}  
\label{eq:r1}
\end{equation}
where (l) denotes the liquid phase (i.e., magma ocean).
The oxygen reacts with atmospheric hydrogen, producing H$_2$O: 
\begin{equation}
\frac{1}{2} \mathrm{O_2} + \mathrm{H_2} \rightleftharpoons  \mathrm{H_2O}, \tag{R2}    
\label{eq:r2}
\end{equation}
while SiO reacts with H$_2$, producing SiH$_4$:
\begin{equation}
\mathrm{SiO} + \mathrm{3H_2} \rightleftharpoons \mathrm{SiH_4} + \mathrm{H_2O}. \tag{R3}    
\label{eq:r3}
\end{equation}

For the chemistry of the atmospheric region that is not in direct contact with the ground, we consider the reactions R2 and R3, but not R1, as the vaporization of SiO$_2$ occurs only at the magma ocean surface.
In addition, we consider the condensation of silicates in the atmosphere through the back reaction of R1, which produces $\mathrm{SiO}_{2}$ condensates;
\begin{equation}
 \mathrm{SiO} + \frac{1}{2} \mathrm{O_2} \rightarrow \mathrm{SiO}_{2} (\mathrm{c}), \tag{R4}
\label{eq:r4} 
\end{equation}
 and another condensation path producing SiO condensates;
\begin{equation}
   \mathrm{SiO} \rightarrow \mathrm{SiO (c)}, \tag{R5}
\label{eq:r5} 
\end{equation}
where (c) denotes condensates in liquid or solid phases.
These condensates are assumed to completelly rain out in the atmosphere, for simplicity. Namely, it is assumed that the atmospheric SiO vapor pressure follows the saturation vapor pressure if the condensation happens. 

The equilibrium constants for these reactions, $K_\mathrm{eq}$, are given by
\begin{eqnarray}
K_\mathrm{eq,R1}&=&	 \frac{\phi_\mathrm{SiO} \phi_\mathrm{O_2}^{1/2}}{a_\mathrm{SiO_2 (l)}} {\left(\frac{P_\mathrm{SiO}}{P_0} \right) \left(\frac{P_\mathrm{O_2}}{P_0} \right)^{1/2} },
\label{eq:k1}
\\
K_\mathrm{eq,R2}&=&\frac{\phi_\mathrm{H_2O}}{\phi_\mathrm{O_2}^{1/2}\phi_\mathrm{H_2}} 
{\frac{(P_\mathrm{H_2O}/P_0)}{(P_\mathrm{O_2}/P_0)^{1/2}(P_\mathrm{H_2}/P_0)}},
\label{eq:k2}
\\
K_\mathrm{eq,R3}&=&\frac{\phi_\mathrm{SiH_4}\phi_\mathrm{H_2O}}{\phi_\mathrm{SiO} {\phi_\mathrm{H_2}^3}}{\frac{(P_\mathrm{SiH_4}/P_0)(P_\mathrm{H_2O}/P_0)}{(P_\mathrm{SiO}/P_0)(P_\mathrm{H_2}/P_0)^3}},
\label{eq:k3}
\\
K_\mathrm{eq,R4}&=& \frac{\phi_\mathrm{SiO} \phi_\mathrm{O_2}{^{1/2}}}{a_\mathrm{SiO_2 (c)}} {\left(\frac{P_\mathrm{SiO}}{P_0} \right) \left(\frac{P_\mathrm{O_2}}{P_0} \right)^{1/2} }, 
\label{eq:k4}
\\
K_\mathrm{eq,R5}&=&\frac{\phi_\mathrm{SiO}}{a_\mathrm{SiO (c)}} {\left(\frac{P_\mathrm{SiO}}{P_0} \right)}, 
\label{eq:k5}
\end{eqnarray}
where $a_\mathrm{SiO_2 (l)}$ is the activity of SiO$_2$ in the magma, $a_\mathrm{SiO_2 (c)}$ and $a_\mathrm{SiO (c)}$ are the activities of $\mathrm{SiO_2}$ and $\mathrm{SiO}$ in the condensates, respectively, $\phi_i$ is the fugacity coefficient of gas species $i${,} $P_i$ is the partial pressure of gas species $i${, and $P_0$ is the reference pressure of the equilibrium constants, which is set at 1~bar. The reference pressure is added to describes the dimensionless form of
the equilibrium constants \citep[][]{Heng+2016}.}
For simplicity, we ignore the non-ideal behavior of all species.
We take $a_\mathrm{SiO_2 (l)}$, $a_\mathrm{SiO_2 (c)}$ and $a_\mathrm{SiO (c)}$ as unity,  assuming that the magma is an ideal, pure SiO$_2$ liquid and that each condensate consists of a ideal pure component. In addition, we take the fugacity coefficients of all gas species to be unity as well, assuming ideal gas behavior due to the lack of public available data on the fugacity coefficients of SiH$_4$ and SiO, to our knowledge.
{Although the fugacity coefficients of some gas species, such as H$_2$ and H$_2$O, are known to deviate from unity at pressures above about 1000~bar \citep[e.g.,][]{Kite+2020, Tian+2024}, we neglect these deviations in this study. This is a reasonable simplification because the ideal gas assumption for all species ensures theoretical consistency in evaluating the chemical potentials and equilibrium constants, even when the fugacity coefficients of some species are unknown. Such an approach has also been adopted in previous models  \citep{Charnoz+2023,Misener+2023}}

The equilibrium constants for R1-R4 are determined by calculating the difference in Gibbs free energy between the reactants and products, while we use the formula of $K_\mathrm{eq,R5}$ determined by \citet{Gail+2013} who experimentally determined the vapor pressure over solid SiO.
All Gibbs free energy values are calculated using data from the JANAF database \citep{Chase1998}. {Because the Gibbs free energy data provided in JANAF are available up to 6000~K for most molecules, the temperature range considered in this study is limited accordingly.}
Note that JANAF provides thermodynamic data of SiO$_2$ (l) up to 4500~K; we extrapolate the data to higher temperatures, as done by \citet{Misener+2023}. We use a least-squares method to derive the following expression fitting the tabulated equilibrium constants at temperatures above 3000 K.
\begin{eqnarray}
\ln K_\mathrm{eq, R1} &=& -\frac{87043.88}{T}+26.5665,
\end{eqnarray}
where the temperature is in Kelvin.
The function yields the square of the correlation function, $R$, minus unity (i.e., $|R^2-1|$) of at most 0.006.
As noted in \citet{Misener+2023}, this produces good agreement with the vapor pressure of SiO$_2$ melt shown in \citet{Visscher+2013}. Although there are slight differences due to different thermodynamic values, we confirmed that the derived vapor pressure of SiO$_2$ melt differs by no more than 60~\% for temperatures from 1500~K to 6000~K.

\subsubsection{Water dissolution}
It is known that H$_2$O 
is highly soluble in
molten silicate
\citep[e.g.,][]{Papale1997}.
Thus, we take this effect into account
by using an empirical fitting formula  
expressed as
\begin{eqnarray}
X_\mathrm{H_2O} = \alpha
{\left( \frac{P_\mathrm{H_2O, gr}}{\tilde{P_0}}\right)^\beta}
\label{eq:diss}
\end{eqnarray}
where $X_\mathrm{H_2O}$ is the mass concentration
of H$_2$O in magma, $P_\mathrm{H_2O,gr}$ is the ground pressure of H$_2$O{, $\tilde{P_0}$ is the unit conversion factor from bar to the target pressure unit (e.g., 
10$^5$~Pa/bar in SI or 
10$^6$~dyn/cm$^2$/bar in cgs)}
, and $\alpha$ and $\beta$ are empirically derived constants{, with units of bar$^{-\beta}$ and dimensionless, respectively.} 
We use the values of $\alpha$ and $\beta$ obtained by \citet{Schaefer+2016}, who
derived this equation by fitting them to the results of \citet{Papale1997}; 
$\alpha=3.44\times 10^{-4.3}${~bar$^{-0.74}$} and $\beta=0.74$.  

{The experimental data used for the calibration of water solubility in \citet{Papale1997} cover a wide range of silicate compositions, including pure SiO$_2$, as well as pressure and temperature conditions ranging from 1~bar to $3 \times 10^4$~bar and from 900~K to 1700~K, respectively~\citep[see Table 3 of][for the details]{Papale1997}.}
 $X_\mathrm{H_2O}$ derived from these $\alpha$ and $\beta$ values is consistent with the solubility of water into pure SiO$_2$,  which was experimentally determined under pressures of  1--9 kbar {and temperatures of 1383--1623~K} by \citet{Kennedy1962} and \citet{Holtz+2000}, with a deviation of less than 60~\%. 
{We discuss the sensitivity of our calculated SiH$_4$ abundance to the solubility law parameter values in Sec.\ref{sec:d_sol}.}

Note that we set the upper limit of $X_\mathrm{H_2O}$ at 10~wt\% since water-rich magma behaves differently from
hydrous silicate magma, becoming fully  miscible
\citep[][]{Newton+2008}.
This change in the behavior of the SiO$_2$--H$_2$O system is also observed in solubility measurements, where the equilibrating pressure increases with dissolved H$_2$O for $X_\mathrm{H_2O}$ below 10~wt\% but remains constant above this threshold \citep[see Fig.~14 of][]{Kennedy1962}. 
In our results, $X_\mathrm{H_2O}$ reaches this upper limit only within a narrow parameter range where 
$P_\mathrm{H_2, gr}\geq10^{4.8}$~bar and $T_\mathrm{gr}\geq5000$~K, as shown in Fig.~\ref{fig:limit}. Although our results within this parameter range may suggest a fully miscible magma ocean, investigating such a state is beyond the scope of this study. 

\subsubsection{Mass balance}
\label{sssec:mb}
The total numbers of oxygen and silicon contained in the atmosphere plus the magma ocean are respectively given by
\begin{equation}
    N^\mathrm{tot}_\mathrm{O} = 
    2N_\mathrm{SiO_2(l)} + N_\mathrm{SiO} + 2 N_\mathrm{O_2} + N_\mathrm{H_2O} + N_\mathrm{H_2O}^\mathrm{(m)}, 
\label{eq:total_O}
\end{equation}
\begin{equation}
    N^\mathrm{tot}_\mathrm{Si} = N_\mathrm{SiO_2(l)} + N_\mathrm{SiO} + N_\mathrm{SiH_4}, 
\label{eq:total_Si}
\end{equation}
where $N_\mathrm{O_2}$, $N_\mathrm{H_2O}$, $N_\mathrm{SiO}$, and $N_\mathrm{SiH_4}$ are the numbers of the gaseous molecules O$_2$, H$_2$O, SiO, and SiH$_4$, respectively, 
$N_\mathrm{SiO_2(l)}$ is that of the SiO$_2$ liquid (or the magma ocean), 
and $N_\mathrm{H_2O}^\mathrm{(m)}$ is that of water dissolved in the magma ocean.
The number of dissolved H$_2$O, $N_\mathrm{H_2O}^\mathrm{(m)}$, is related to the mass of the core, $M_\mathrm{core}$, as
\begin{equation}
    N_\mathrm{H_2O}^\mathrm{(m)} = 
    \frac{m_\mathrm{H_2O}N_\mathrm{H_2O}^\mathrm{(m)}}{m_\mathrm{SiO_2}N_\mathrm{SiO_2(l)}} \cdot \frac{m_\mathrm{SiO_2}N_\mathrm{SiO_2(l)}}{m_\mathrm{H_2O}} 
    = X_\mathrm{H_2O} \frac{\zeta M_\mathrm{core}}{m_\mathrm{H_2O}},
\end{equation}
where $M_\mathrm{core}$ is the mass of the interior core, $\zeta$ is the mass fraction of the well-mixed magma ocean in the interior core and $X_\mathrm{H_2O}$ is given by ${m_\mathrm{H_2O}N_\mathrm{H_2O}^\mathrm{(m)}}/{m_\mathrm{SiO_2}N_\mathrm{SiO_2(l)}}$.

{In order to isolate the contributions of species derived from the vaporization of SiO$_2$, we}
 introduce the following new numbers
\begin{eqnarray}
    N_\mathrm{O} &\equiv& N^\mathrm{tot}_\mathrm{O} - 2N_\mathrm{SiO_2(l)} - \mathcal{N}_\mathrm{O}, \\
    N_\mathrm{Si} &\equiv& N^\mathrm{tot}_\mathrm{Si} - N_\mathrm{SiO_2(l)} - \mathcal{N}_\mathrm{Si}, 
\end{eqnarray}
where $\mathcal{N}_\mathrm{O}$ and $\mathcal{N}_\mathrm{Si}$ are the numbers of exotic oxygen and silicon atoms, respectively, originating from reservoirs other than the vaporization of SiO$_2$ on the magma ocean. {For example, protoplanetary nebulae contain such exotic oxygen in the form of molecules such as H$_2$O and CO, which would be accreted onto planets during their formation. In addition, such volatiles may also be introduced through degassing from the planetary interior, independently of the vaporization of SiO$_2$.}
Accreted nebula gases and outgassed volatile species can {therefore} be considered as potential reservoirs for such exotic molecules.
For simplicity, we neglect exotic molecules in our simulations; however, we discuss later their potential influence, along with other molecules that could affect the SiH$_4$ abundance, in Section ~\ref{sec:d_oc}.

In hydrostatic equilibrium, the base of the atmosphere has the highest density and contains the majority of the gaseous material, implying that the elemental ratios of the entire atmosphere can be approximately represented by the partial pressure ratios at the surface. Although this approximation is not strictly accurate in cases where an atmosphere is not uniformly mixed, as considered in this study, we adopt it to reduce computational iterations and costs following previous studies \citep{Charnoz+2023,Misener+2023}.
Under this assumption and an assumption that $\mathcal{N}_\mathrm{O}$ and $\mathcal{N}_\mathrm{Si}$ are negligible, we rewrite Eqs.~(\ref{eq:total_O}) and (\ref{eq:total_Si}) as
\begin{equation}
    \frac{N_\mathrm{O}}{N_\mathrm{H_2}}
    = 
    \frac{P_\mathrm{SiO, gr}}{P_\mathrm{H_2, gr}} + 
    2\frac{P_\mathrm{O_2, gr}}{P_\mathrm{H_2, gr}} + 
    \frac{P_\mathrm{H_2O, gr}}{P_\mathrm{H_2, gr}}+ 
    \frac{X_\mathrm{H_2O}}{9} \frac{\zeta M_\mathrm{core}}{M_\mathrm{H_2}},
\label{eq:total_O_2}
\end{equation}
\begin{equation}
    \frac{N_\mathrm{Si}}{N_\mathrm{H_2}} = \frac{P_\mathrm{SiO, gr}}{P_\mathrm{H_2, gr}} + \frac{P_\mathrm{SiH_4, gr}}{P_\mathrm{H_2, gr}},
\label{eq:total_Si_2}
\end{equation}
where the subscription of $P$, gr, denotes their partial pressure at the ground and $M_\mathrm{H_2}$ is the total mass of H$_2$ in the atmosphere.
Since all the oxygen and silicon except the exotic species (i.e., $\mathcal{N}_\mathrm{O}$ and $\mathcal{N}_\mathrm{Si}$) come from SiO$_2(l)$, the relation 
\begin{equation}
    N_\mathrm{O} = 2 N_\mathrm{Si}
\label{eq:O2Si}
\end{equation}
holds. Thus, from Eqs.(\ref{eq:total_O_2})-(\ref{eq:O2Si}), 
we obtain
\begin{eqnarray}
X_\mathrm{H_2O} &=& \left(
        2 \frac{P_\mathrm{SiH_4, gr}}{P_\mathrm{H_2, gr}} + 
          \frac{P_\mathrm{SiO, gr}}{P_\mathrm{H_2, gr}} - 
          \frac{P_\mathrm{H_2O, gr}}{P_\mathrm{H_2, gr}} - 
        2 \frac{P_\mathrm{O_2, gr}}{P_\mathrm{H_2, gr}}
\right) \notag \\
  && \times \frac{9M_\mathrm{H_2}}{\zeta M_\mathrm{core}},
\label{eq:mass_balance}
\end{eqnarray}
$M_\mathrm{H_2}$ is given by 
\begin{equation}
    M_\mathrm{H_2} = M_\mathrm{atm} y_\mathrm{H_2},
\label{eq:M_H2}
\end{equation}
where $M_\mathrm{atm}$ is the total atmospheric mass and $y_\mathrm{H_2}$ is the $\mathrm{H_2}$ mass fraction in the atmospheres.
We should note that the assumption of $N_\mathrm{O}/N_\mathrm{Si}=2$ (Eq.~\ref{eq:O2Si}) becomes invalid if oxygen reservoirs such as Fe-oxides, rather than SiO$_2$, dominate in the magma ocean. In contrast, the assumption is likely valid for reduced magma ocean (i.e., FeO-free), which is of interest in this study. We later discuss that the redox state of magma ocean in our model encompasses that of enstatite chondrites in Section~\ref{sec:d_r}.

\subsection{Atmospheric structure}
\subsubsection{Atmospheric composition at ground}
Firstly, we find the ratio of the partial pressures of species $i$ to that of H$_2$ at ground in equilibrium: 
\begin{equation}
x_i \equiv P_{i,gr}/P_\mathrm{H_2,gr}, 
\label{eq:xi}
\end{equation}
 for $i$ = SiH$_4$, SiO, O$_2$, and H$_2$O. 
From the equilibrium conditions given by Eqs.~(\ref{eq:k1})-(\ref{eq:k3}), one obtains $x_i$ as a function of $x_\mathrm{H_2O}$ for given $P_\mathrm{H_2,gr}$ and $T_\mathrm{gr}$ as
\begin{eqnarray}
    x_\mathrm{SiO}   &=& K_\mathrm{eq,R1}K_\mathrm{eq,R2} {\left(\frac{P_\mathrm{H_2, gr}}{P_0} \right)^{-1}}  \, x_\mathrm{H_2O}^{-1},
    \label{eq:x_sio}
    \\
    x_\mathrm{O_2}   &=& K_\mathrm{eq,R2}^{-2} {\left(\frac{P_\mathrm{H_2, gr}}{P_0} \right)^{-1}}  \, x_\mathrm{H_2O}^2,
        \label{eq:x_o2}
    \\
    x_\mathrm{SiH_4} &=& K_\mathrm{eq,R1}K_\mathrm{eq,R2}K_\mathrm{eq,R3} {\left(\frac{P_\mathrm{H_2, gr}}{P_0} \right)}  \, x_\mathrm{H_2O}^{-2},
    \label{eq:x_sih4}
\end{eqnarray}
which are substituted into Eq.~(\ref{eq:mass_balance}) to yield
\begin{eqnarray}
    X_\mathrm{H_2O} &=& 
    \frac{9M_\mathrm{H_2}}{\zeta M_\mathrm{core}} 
    (2x_\mathrm{SiH_4}+x_\mathrm{SiO}-x_\mathrm{H_2O}-2x_\mathrm{O_2})  \nonumber
    \\
    &\equiv&     \frac{9M_\mathrm{H_2}}{\zeta M_\mathrm{core}} f (x_\mathrm{H_2O}, P_\mathrm{H_2,gr}, T_\mathrm{gr}).
    \label{eq:f}
\end{eqnarray}
Meanwhile, the dissolution equilibrium given by Eq.~(\ref{eq:diss}) can be expressed as
\begin{equation}
    X_\mathrm{H_2O} = \alpha 
    {\left( \frac{P_\mathrm{H_2, gr}}{\tilde{P_0}}\right)^\beta}x_\mathrm{H_2O}^\beta \equiv g (x_\mathrm{H_2O}, P_\mathrm{H_2,gr}).
       \label{eq:g} 
\end{equation}
Thus, we solve 
\begin{equation}
    \frac{9M_\mathrm{H_2}}{\zeta M_\mathrm{core}} f (x_\mathrm{H_2O}, P_\mathrm{H_2, gr}, T_\mathrm{gr})
    = g (x_\mathrm{H_2O}, P_\mathrm{H_2, gr}).
    \label{eq:feqg}
\end{equation} 
For $\zeta$, we adopt a value of 0.5, assuming that the rocky part of an interior core is mostly molten.
Such conditions are more likely to be satisfied for planets with higher ground temperature. For example. it  could be reasonable for $T_\mathrm{gr}\geq 3000$~K if the magma adiabats and the melting curve of the rock were same with those of chondritic mantle \citep[see Fig.~3 of][]{Kite+2020}.
We discuss how the 
magma fraction affects on our calculations in Section~\ref{sec:d_t}.

Using the relation between the ground pressure and total mass of the atmosphere, 
$P_\mathrm{gr} \approx \left({GM_\mathrm{core}}/{4 \pi R_\mathrm{core}^4}\right) M_\mathrm{atm}$
and Eq.~(\ref{eq:M_H2}), one can relate $M_\mathrm{H_2}$ to $P_\mathrm{gr}$ as 
\begin{eqnarray}
    M_\mathrm{H_2} &=& \frac{4 \pi R_\mathrm{core}^4}{GM_\mathrm{core}}  P_\mathrm{gr} 
    \left(\frac{1}{1-Y_\mathrm{He}}+\Sigma_i x_i \frac{m_i}{m_\mathrm{H_2}}
    \right)^{-1},
\end{eqnarray}
where $m_i$ is the molecule mass of species $i$.
Also, the ground pressure is related to the H$_2$ partial pressure as 
\begin{equation}
    P_\mathrm{gr} = (\chi_\mathrm{H_2}^{-1} + \Sigma_i x_i) P_\mathrm{H_2, gr}.
        \label{eq:Mh2}
\end{equation}
From the above three equations with given $T_\mathrm{gr}$ and $P_\mathrm{H_2, gr}$, we obtain $x_\mathrm{H_2O}$ and thereby the other $x_i$. 
Then, the molar fraction of atmospheric gas species, $x'$, at ground can be obtained as
\begin{eqnarray}
     x'_i&=&  x_i \frac{P_\mathrm{H_2, gr}}{P_\mathrm{gr}}, \label{eq:x'x} \\
     x'_\mathrm{H_2+He}&=&  \frac{1}{\chi_\mathrm{H_2}}\frac{P_\mathrm{H_2, gr}}{P_\mathrm{gr}}. 
     \label{eq:xh2he}
\end{eqnarray}
{In Section \ref{sec:res}, we present the molar fraction, $x_i'$, but not $x_i$.}

{Hence, there are 15 unknown variables: $x_\mathrm{i}'$, $x_\mathrm{i}$ and $P_\mathrm{i,gr}$ for each of SiH$_4$, SiO, O$_2$, and H$_2$O, along with $P_\mathrm{gr}$, $M_\mathrm{H_2}$ and $x_\mathrm{H_2+He}'$. These variables are solved using 15 equations (Eqs.~\ref{eq:xi}--\ref{eq:x_sih4} and Eqs.~\ref{eq:feqg}--\ref{eq:xh2he}), given parameters of $P_\mathrm{H_2, gr}$ and $T_\mathrm{gr}$, and assuming values for $\zeta$, $M_\mathrm{core}$, $R_\mathrm{core}$, $\alpha$, and $\beta$ at the ground. }

\subsubsection{Atmospheric structure above ground}
 From these values at the ground, we vertically integrate a hydrostatic equation with an adiabatic lapse rate for $P > P_\mathrm{rcb}$ and an isothermal structure for $P\leq P_\mathrm{rcb}$ to calculate
 the temperature-pressure-density profile from the ground to the optical {photospheric} radius. We account for the altitude dependence of planetary gravity and the adiabatic lapse rate when integrating the atmospheric structure, while ignoring the self-gravity of the atmosphere, the latent heat of condensation and thermal conduction.  The adiabatic lapse rate is calculated with the additive volume law, using the fractions of all gas species at each altitude, the equation of
state for an H-He mixture \citep{Chabrier+2021} and the heat capacity of SiH$_4$, SiO, O$_2$, and H$_2$O from the JANAF database \citep{Chase1998}.
 For this vertical integration, we use the explicit Runge-Kutta method with fourth-order accuracy. 
The simplified thermal structure is discussed as one of the caveats of our model in Sec~\ref{sec:d_tp}. 

In addition, we calculate the vertical profile of the atmospheric composition in chemical equilibrium among the reactions R2-R5, simultaneously. To account for the rainout of condensates in our chemical equilibrium calculation, we calculate the chemical equilibrium composition at each altitude, considering the effective abundances of elements remaining in the gas phase at the one grid below. This scheme is based on the chemical equilibrium calculation method implemented in an open-source code FASTCHEM COND \citep{Kitzmann+2024}.

 To obtain the chemical equilibrium composition above the ground, we perform the Gibbs free energy minimization calculations with a Newton-Raphson method.
 The calculation is performed for Eqs.(\ref{eq:k2}) - (\ref{eq:k3}), if $P_\mathrm{SiO}$ is smaller than its saturated vapor pressure determined by Eqs.(\ref{eq:k4}) or (\ref{eq:k5}).
If $P_\mathrm{SiO}$ is super-saturated, we perform the calculations including the Eqs.(\ref{eq:k4}) or/and (\ref{eq:k5}) to account for condensation until $P_\mathrm{SiO}$ reaches the saturated value.
As the initial guess of the calculations, we use $x'_i$ at the one grid below for the non-saturated SiO system, while we use the equilibrium value of $x'_i$ derived from the condensates-free chemistry (i.e., R2 and R3) at the same grid for the super-saturated one. 
In cases that $P_\mathrm{SiO}$ is super-saturated to condense both SiO(c) and SiO$_2$(c), the initial guess from the condensates-free chemistry sometimes did not work, numerically. This is likely due to the competition between the condensates SiO(c) and SiO$_2$ (c) for the elements Si and O. The numerical problem in such cases is also discussed in \citet{Kitzmann+2024}.
To avoid this numerical problem, we choose an initial guess derived from the chemistry including only a single condensate, eigher SiO or SiO$_2$, which numerically works with iterating the initial guess.
Note that such chemical equilibrium calculations involving multiple condensates 
of the same elements might depend on the initial guess. We discuss the problem of the multiple condensates system in Section~\ref{sec:d_c}.

\section{Result} \label{sec:res}
\subsection{Impact of water dissolution on atmosphere}
\begin{figure}
 \begin{minipage}{0.5\textwidth}
    \begin{center}
        ($a$) with water dissolution
  \includegraphics[width=\columnwidth]{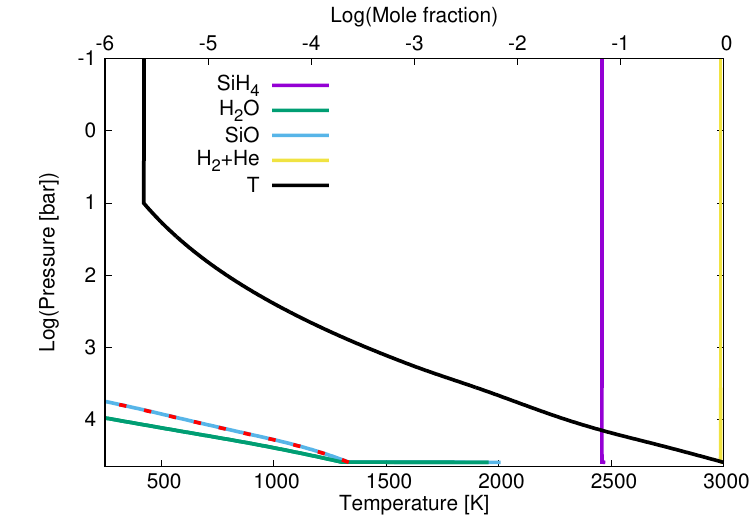}
   \end{center}
 \end{minipage}
   \begin{minipage}{0.5\textwidth}
    \begin{center}
        ($b$) without water dissolution
  \includegraphics[width=\textwidth]{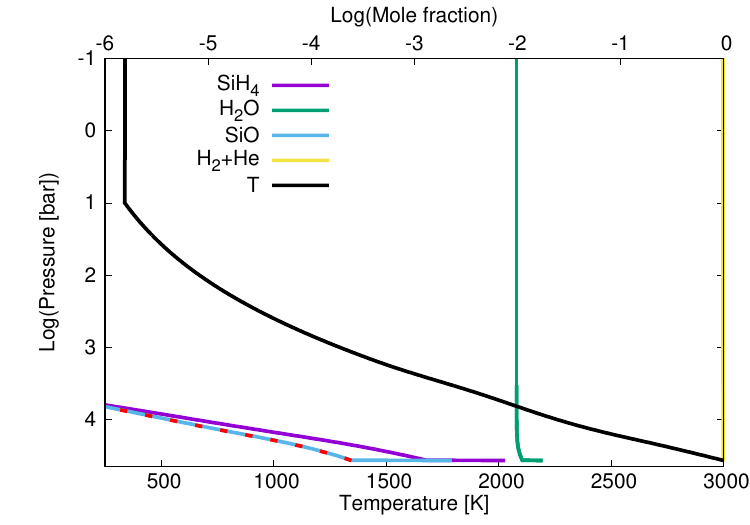}
   \end{center}
 \end{minipage}
\caption{
Atmospheric structure 
with water dissolution into magma ocean ($a$) and without it ($b$)
for a tropopause pressure of 10~bar, a ground H$_2$ pressure of 3$\times10^4$~bar and a ground temperature of 3000~K. 
The vertical distributions of SiH$_4$ (purple),  H$_2$O (green), SiO (cyan) and the sum of H$_2$ and He (orange), and  temperature  (black) are shown 
as functions of
pressure from the ground pressure to 0.1 bar. The red dotted lines represent the saturated vapor pressure of SiO determined by the reaction (R5).
}
\label{fig:atom_comp1}
\end{figure}

First, we demonstrate the impact of water dissolution into magma on the atmospheric SiH$_4$ abundance.
Figure~\ref{fig:atom_comp1} shows the vertical profiles of atmospheric temperature and composition for given $P_\mathrm{rcb}$, $P_\mathrm{H_2,gr}$ and $T_\mathrm{gr}$ of 10~bar, 3$\times10^4$~bar and 3000~K, respectively, as examples of our calculations.  Each panel of Fig.~~\ref{fig:atom_comp1} shows the atmospheric structure obtained with our model considering the water dissolution ($a$) and with that ignoring it ($b$). 
As shown in Fig.~\ref{fig:atom_comp1}$a$,  SiH$_4$ (purple) is the third most abundant gas, with a molar fraction of about 7~\% throughout the atmosphere, next to H$_2$ and He. 
On the other hand, in the case without the water dissolution (Fig.~\ref{fig:atom_comp1}$b$), the SiH$_4$ fraction is about 0.8~\% at the ground and less than 10$^{-5}$ in pressure lower than 10$^4$~bar.

The difference in the SiH$_4$ abundance is caused by the difference of H$_2$O abundance in the atmosphere due to water dissolution. {This is a natural outcome explained by Le Chatelier’s principle, as the reaction producing SiH$_4$ (R3) is promoted by not only higher abundances of SiO and H$_2$, but also by a lower abundance of H$_2$O.} In Fig.~\ref{fig:atom_comp1},  
the water dissolution into magma reduces the H$_2$O molar fraction (green) at the ground from $\sim1.7$~\% to $\sim0.5$~\%, 
which in turn enhances the SiH$_4$ 
 fraction at the ground by about a factor of 10.
 The dependence of the SiH$_4$ fraction on the H$_2$O fraction can be explained from {$x_\mathrm{SiH_4}' \sim x_\mathrm{SiH_4}$ in Eq.(\ref{eq:x'x}) for this case and} $x_\mathrm{SiH_4} \propto x_\mathrm{H_2O}^{-2}$ in Eq.(\ref{eq:x_sih4}).
Under the two relationships of $x_\mathrm{SiH_4} \propto x_\mathrm{H_2O}^{-2}$ and $\mathcal{N}_\mathrm{O}$/$\mathcal{N}_\mathrm{Si}=2$, SiH$_4$ becomes more abundant than H$_2$O if the atmospheric H$_2$O abundance decreases
by less than $1/2^{1/3}$($\sim0.8$) due to water dissolution 
and if SiO is not a main Si-bearing species.
Therefore, water dissolution into magma enhances the SiH$_4$ abundance and can easily make it the third most abundant species under the reduced condition assumed here.
 
In Fig.~\ref{fig:atom_comp1}$a$, H$_2$O and SiO (cyan) are less abundant than SiH$_4$, having the molar fractions of  0.1~\% levels at the ground. Their fractions decrease with the decrease of pressure and temperature in the atmosphere. 
This is because the SiO condensation (R5) reduces the SiO abundance to its saturated value (red dotted) and the H$_2$O reacting with the more abundant SiH$_4$ (R3) is also reduced in response to the saturated SiO, as explained by Le Chatelier's principle. Thus, SiH$_4$ remains abundant throughout the atmosphere while H$_2$O and SiO do not persist due to the condensation of SiO. 
Conversely, if the SiH$_4$ abundance at the ground is lower than that of H$_2$O, SiH$_4$ does not persist in the low-pressure region of the atmosphere, as shown in our calculation
without water dissolution (Fig.~\ref{fig:atom_comp1}$b$).

We note the jumps in the molar fractions between the ground and the adjacent grid above. The jumps come from the difference between SiO abundance determined by the vaporization of SiO$_2$ liquid (R1) and that determined by the SiO condensation (R5), the latter of which is not considered at the ground in our model. If we incorporated eddy diffusion into our model, the  discontinuities would disappear, but a boundary layer would form near the ground.
A case in which SiO condensation (R5) occurs at the ground corresponds to a magma ocean composed of SiO and SiO$_2$, which is not the focus of this study.

\begin{figure}[t]
 \begin{minipage}{0.5\textwidth}
    \begin{center}    \includegraphics[width=\textwidth]{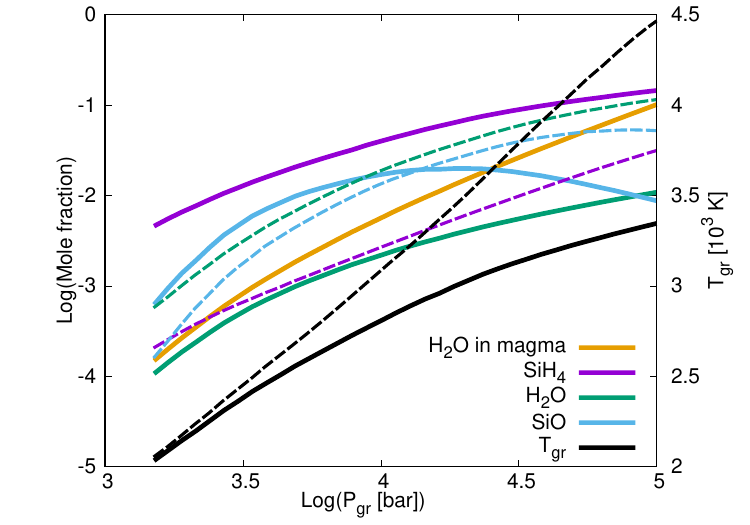}
   \end{center}
 \end{minipage}
\caption{
Molar fractions of 
H$_2$O in the magma ocean ($=X_\mathrm{H_2O}\times 60/18$) (orange),  
atmospheric SiH$_4$ (purple), H$_2$O (green), and SiO (cyan) at the ground, and the ground temperature, $T_\mathrm{gr}$ (black),  are shown as functions of the ground pressure, $P_\mathrm{gr}$, for a resultant tropopause temperature, $T_\mathrm{rcb}=500$~K, and  $P_\mathrm{rcb}=10$~bar. For comparison, the dashed lines represent the results obtained when H$_2$O dissolution into the magma ocean is ignored.
}
\label{fig:atom_comp2}
\end{figure}

Figure~\ref{fig:atom_comp2} shows the atmospheric composition at the ground under various ground pressures, $P_\mathrm{gr}$, 
for a case where the resultant tropopause temperature,  $T_\mathrm{rcb}$, is $500$~K and $P_\mathrm{rcb}=10$~bar.
The molar fraction of SiH$_4$ (purple) increases with $P_\mathrm{gr}$. 
The H$_2$O fraction (green) and the dissolved water fraction in the magma ocean (orange) also increase with $P_\mathrm{gr}$.
This is because higher ground pressure provides the higher ground temperature (black), resulting in more Si and O atoms vaporizing from the magma ocean.
The SiO fraction (cyan) increases with $P_\mathrm{gr}$ at $P_\mathrm{gr}\lesssim10^{4}$~bar
for the same reason but slightly decreases with $P_\mathrm{gr}$ at $P_\mathrm{gr}\gtrsim 10^{4}$~bar.
The slight decrease in the SiO fraction with $P_\mathrm{gr}$ mainly comes from the increase of $P_\mathrm{H_2,gr}$ as the ratio of SiO to SiH$_4$ is proportional to $P_\mathrm{H_2,gr}^{-2}x_\mathrm{H_2O}$ from Eqs.~(\ref{eq:x_sio}) and (\ref{eq:x_sih4}).

\begin{figure}[t]
 \begin{minipage}{0.5\textwidth}
    \begin{center}
\includegraphics[width=\textwidth]{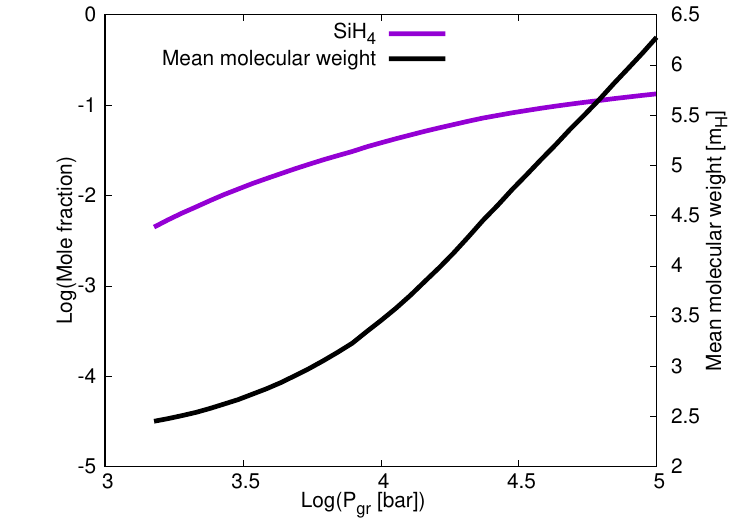}
   \end{center}
 \end{minipage}
\caption{Molar fraction of SiH$_4$ at 0.1~bar  (purple) and atmospheric mean molecular weight at 0.1~bar (black) are shown as functions of the ground pressure, $P_\mathrm{gr}$ for $T_\mathrm{rcb}=500$~K and $P_\mathrm{rcb}=10$~bar. The mean molecular weight is shown in units of atomic hydrogen mass. Note that the mole fractions of SiO and H$_2$O are not shown, as they do not persist at 0.1~bar.}
\label{fig:atom_comp3}
\end{figure}

Figure~\ref{fig:atom_comp3} shows the mole fractions of SiH$_4$ at 0.1~bar (purple) and atmospheric mean molecular weight at 0.1~bar (black) for the atmospheres shown in Fig.~\ref{fig:atom_comp2}. The SiH$_4$ fractions in the upper atmospheric regions remain nearly unchanged from those at the ground (Fig.~\ref{fig:atom_comp2}), increasing from 0.3 to 10~\% with $P_\mathrm{gr}$ from 10$^{3.2}$~bar to 10$^{5}$~bar.
As shown in Figure~\ref{fig:atom_comp2}, the molar fraction of SiH$_4$ is higher than that of H$_2$O at the ground. Under these conditions, the condensation of SiO can completely remove  H$_2$O but cannot eliminate the abundant SiH$_4$ in the upper atmospheric regions, which are colder than the ground. As a result, SiH$_4$ remains the third most abundant species even in the upper atmosphere, as shown in Figure~\ref{fig:atom_comp3}. Our results, presented here and in Fig.~\ref{fig:atom_comp1}$a$, differ significantly from previous models that neglect water dissolution, in which SiH$_4$ is depleted in the upper atmosphere \citep{Misener+2023,Falco+2024}.

As SiH$_4$ is 16 times heavier than H$_2$, its high abundance enhances the atmospheric mean molecular weight from 2.4~$m_\mathrm{H}$ to 6.2~$m_\mathrm{H}$, where $m_\mathrm{H}$ is the mass of an atomic hydrogen, for $P_\mathrm{gr}=10^{3.2}$--10$^{5}$~bar, as shown in Figure~\ref{fig:atom_comp3}. The increase in atmospheric mean molecular weight reduces the atmospheric scale height, thereby influencing the observable properties of sub-Neptunes, such as their optical radii and transmission spectra.

\subsection{Atmospheres with different surface conditions}

\begin{figure*}
  \begin{minipage}{0.5\textwidth}
    \begin{center}   
    ($a$) SiH$_4$ at 0.1 bar
\includegraphics[width=\textwidth]{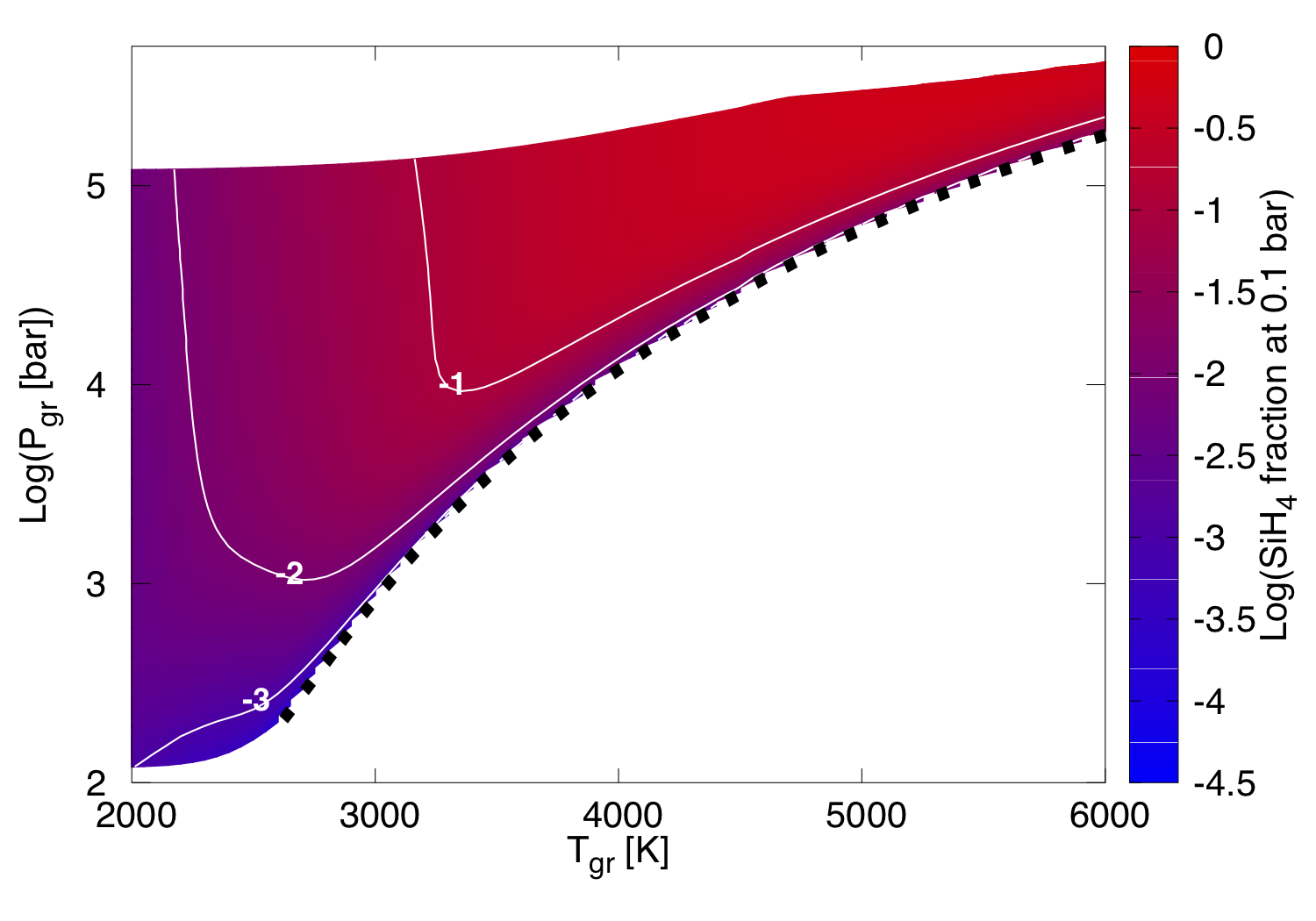}
   \end{center}
 \end{minipage}
   \begin{minipage}{0.5\textwidth}
    \begin{center}   
   ($b$) SiH$_4$ at ground
\includegraphics[width=\textwidth]{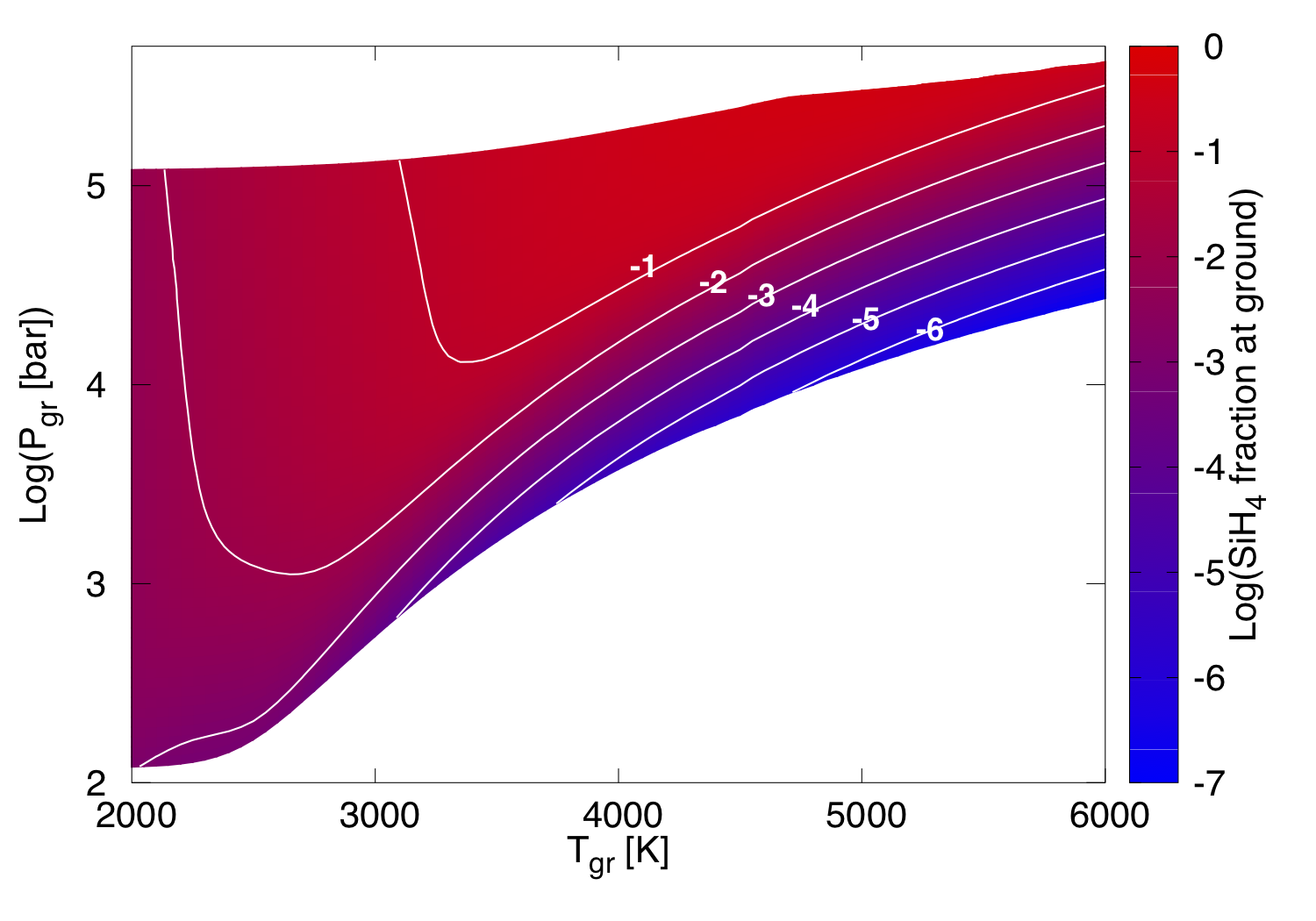}
   \end{center}
 \end{minipage}
   \begin{minipage}{0.5\textwidth}
    \begin{center}   
   ($c$) H$_2$O at ground
\includegraphics[width=\textwidth]{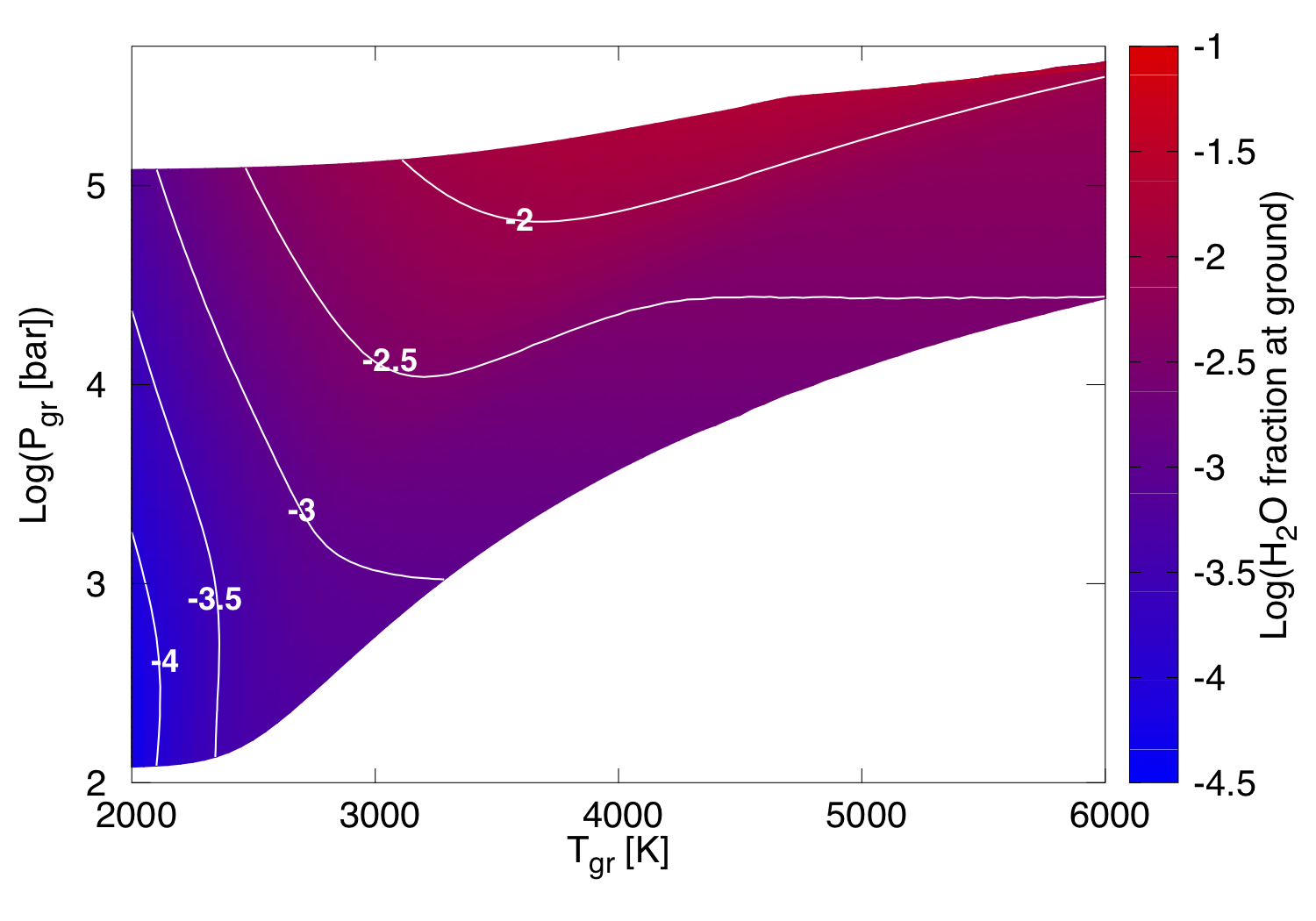}
   \end{center}
 \end{minipage}
   \begin{minipage}{0.5\textwidth}
    \begin{center}   
  ($d$)  SiO at ground
\includegraphics[width=\textwidth]{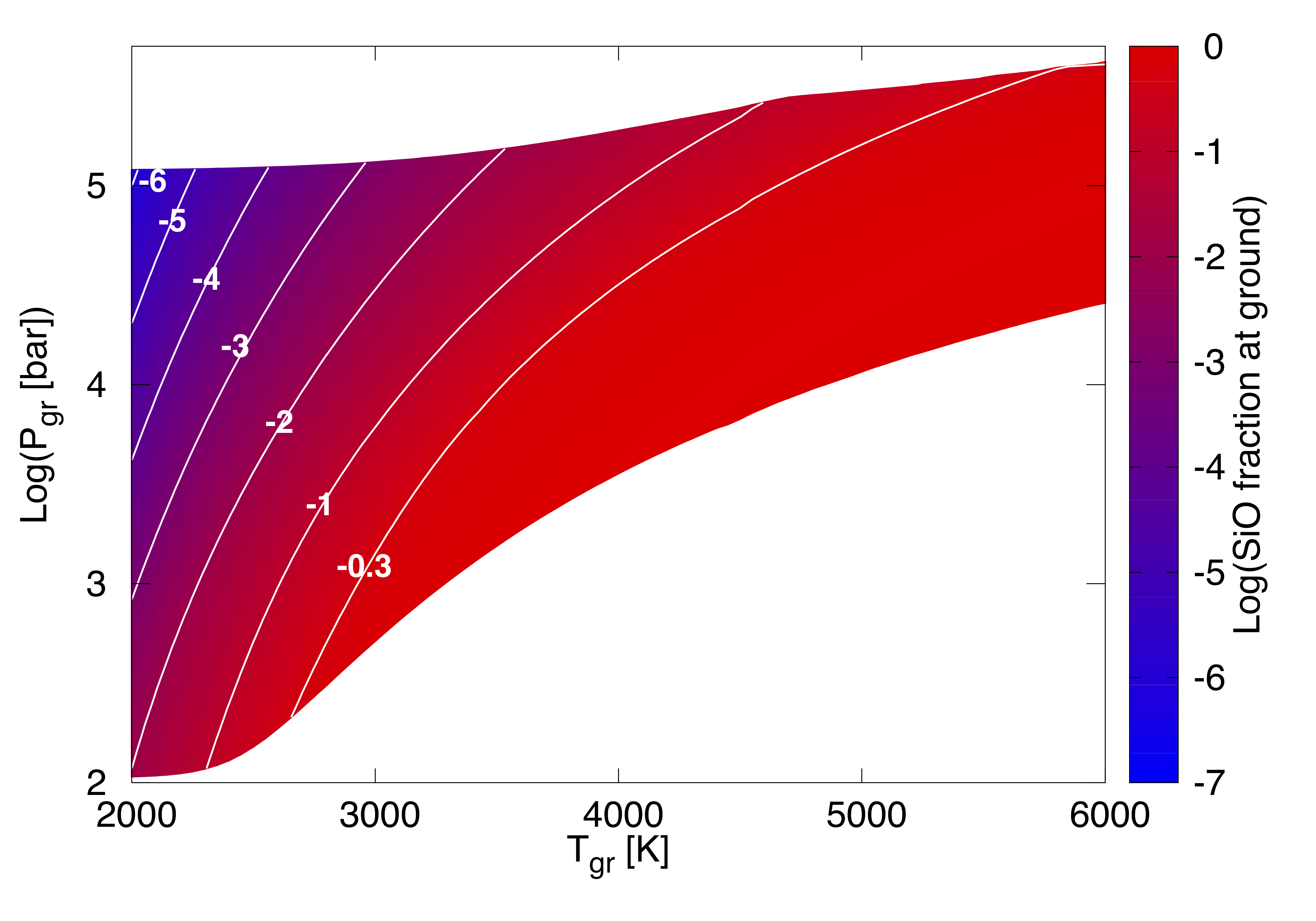}
   \end{center}
 \end{minipage}
  \begin{minipage}{0.5\textwidth}
    \begin{center}   
  ($e$)  H$_2$O concentration in magma
\includegraphics[width=\textwidth]{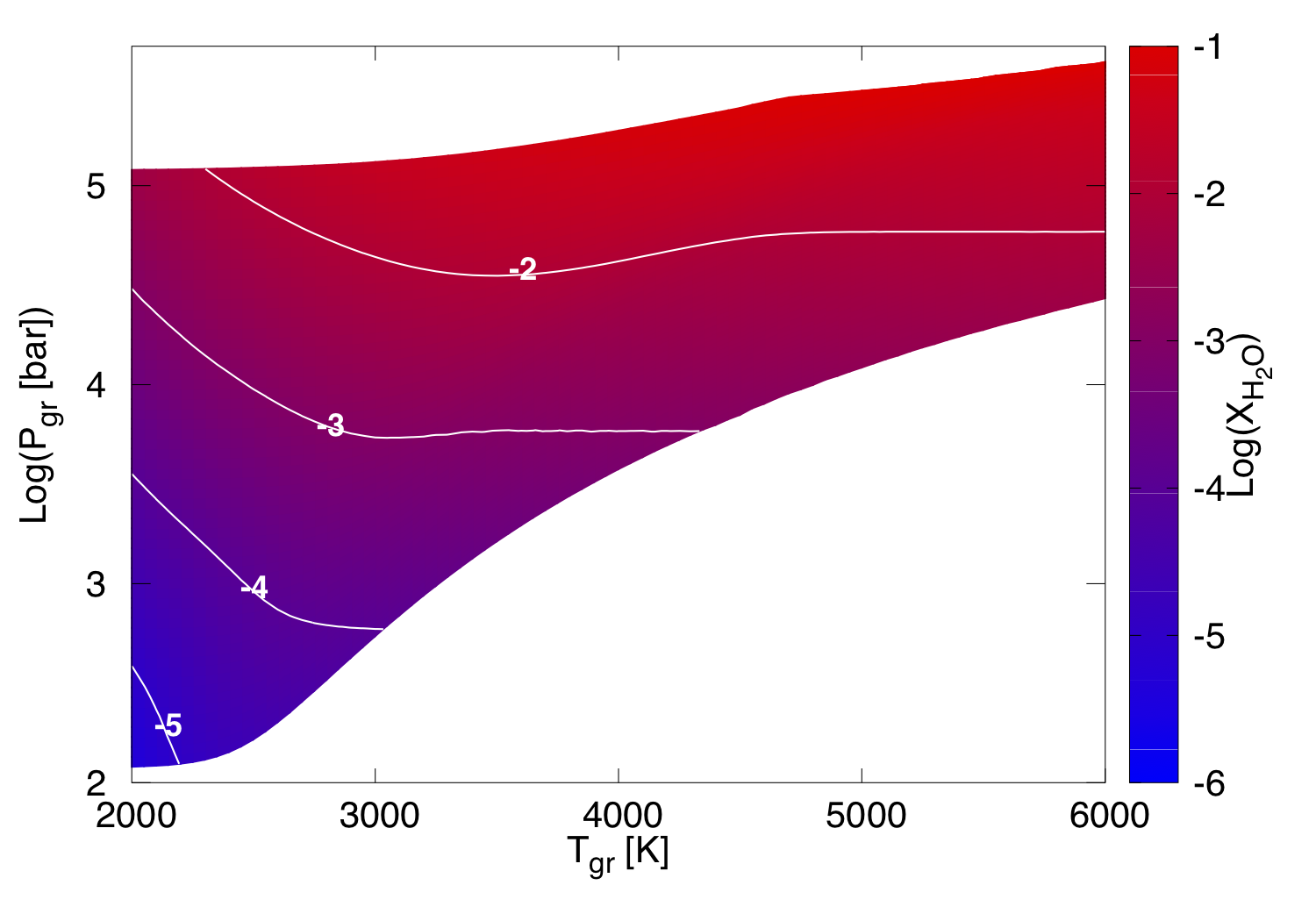}
   \end{center}
 \end{minipage}
 \begin{minipage}{0.5\textwidth}
    \begin{center}   
   ($f$) H$_2$ ground pressure
\includegraphics[width=\textwidth]{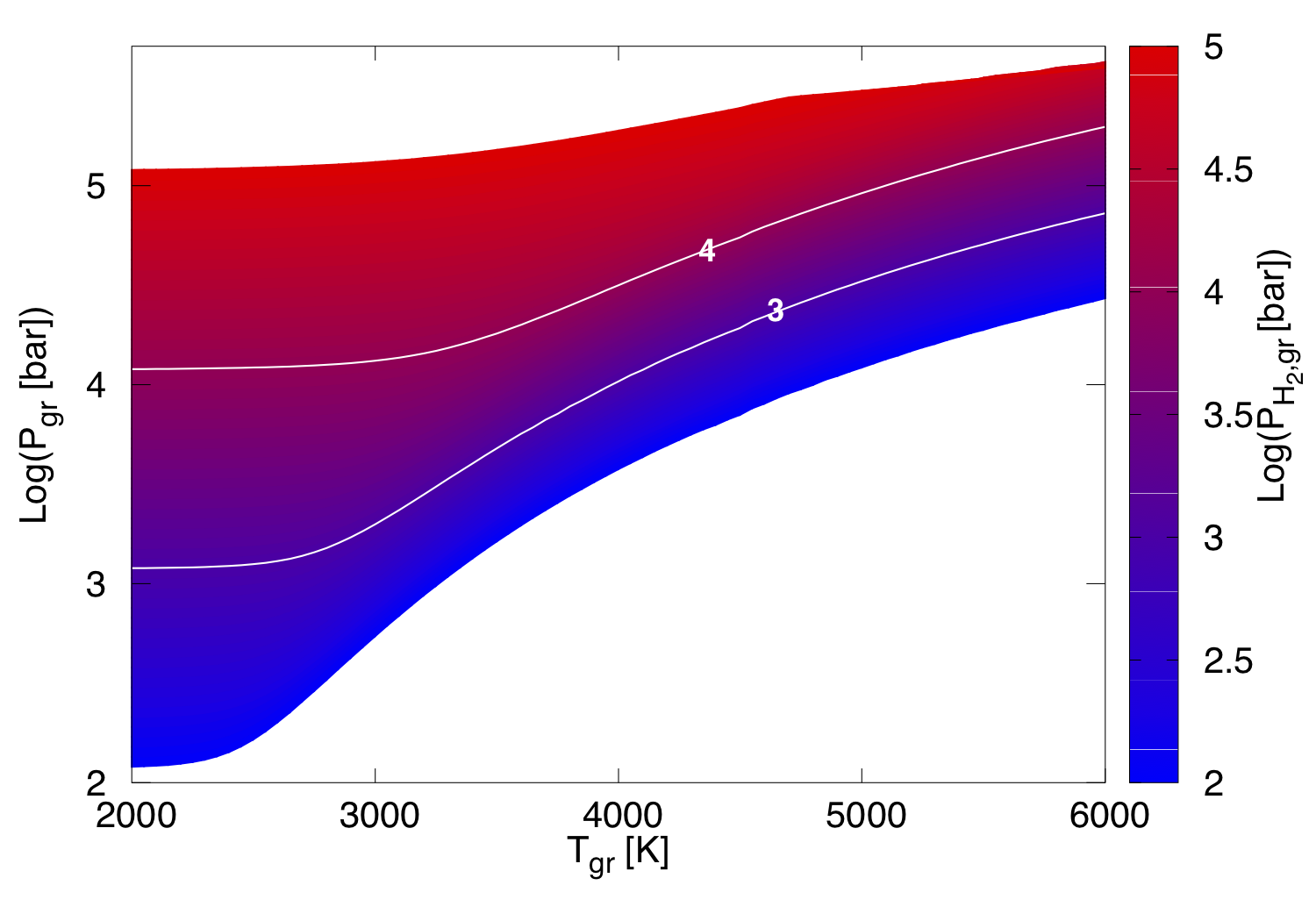}
   \end{center}
 \end{minipage}

\caption{
Atmospheric molar fractions of SiH$_4$ at 0.1~bar ($a$) and at grand ($b$), that of H$_2$O ($c$) and SiO ($d$) at ground, H$_2$O mass concentration in magma ($e$), and H$_2$ ground pressure ($f$) for different ground pressure, $P_\mathrm{gr}$, and temperature, $T_\mathrm{gr}$. Contour counters and lines shows the log10 values of the fractions ($a$--$e$) and the pressure in bar ($f$). Dotted line in panel $a$ represents the condition with a Si/O ratio of 1 at the ground. Note that white regions in panels ($b$--$f$)  represent areas outside the calculated parameter space while those in panel ($a$) indicate the same area as well as parameter spaces where no SiH$_4$ is present at 0.1~bar.
}
\label{fig:atom_compx}
\end{figure*}

\begin{figure*}
 \begin{minipage}{1.0\textwidth}
    \begin{center}   
   ($a$) $P_\mathrm{rcb}=100$~bar \\
\includegraphics[width=0.495\textwidth]{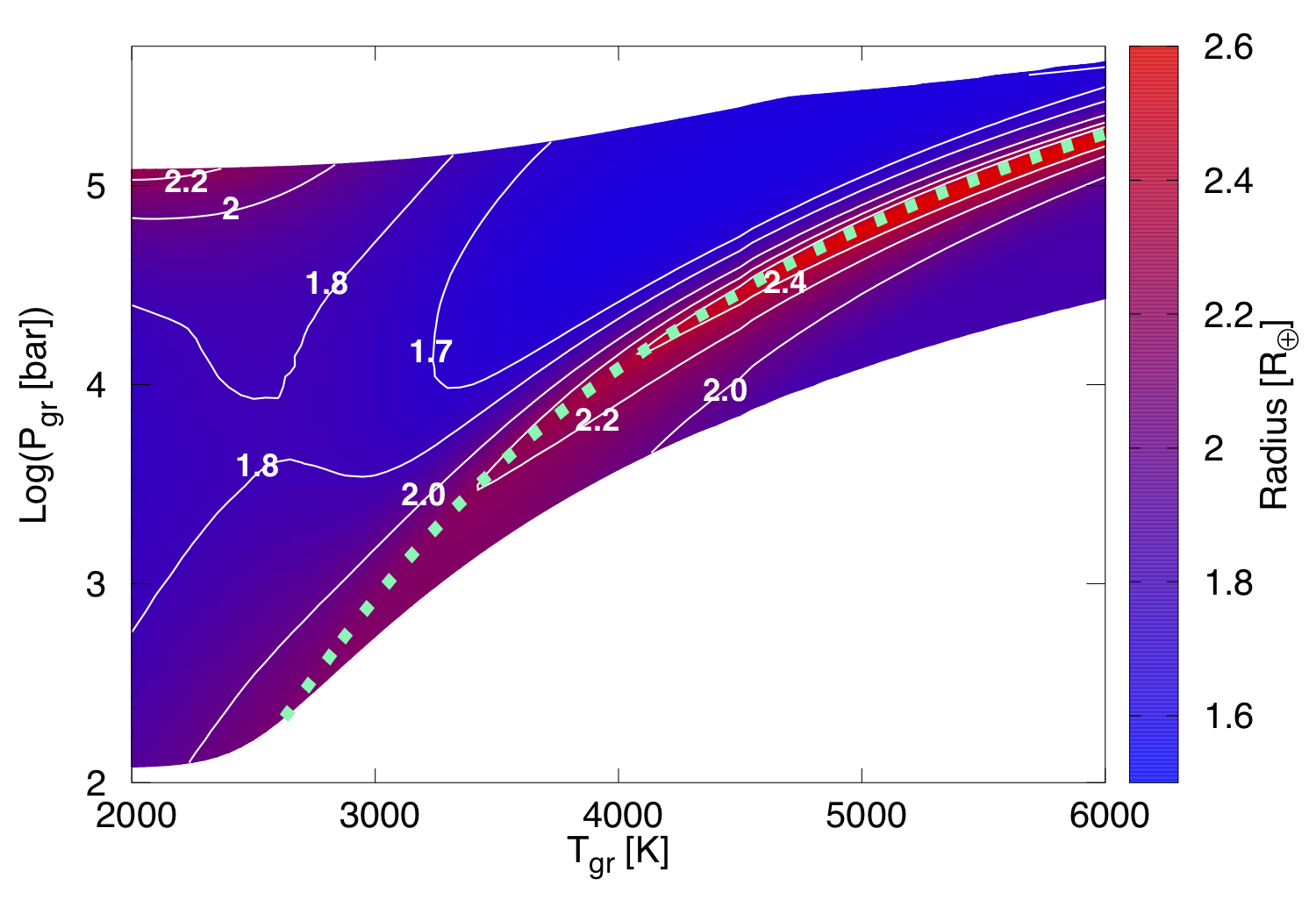}
\includegraphics[width=0.495\textwidth]{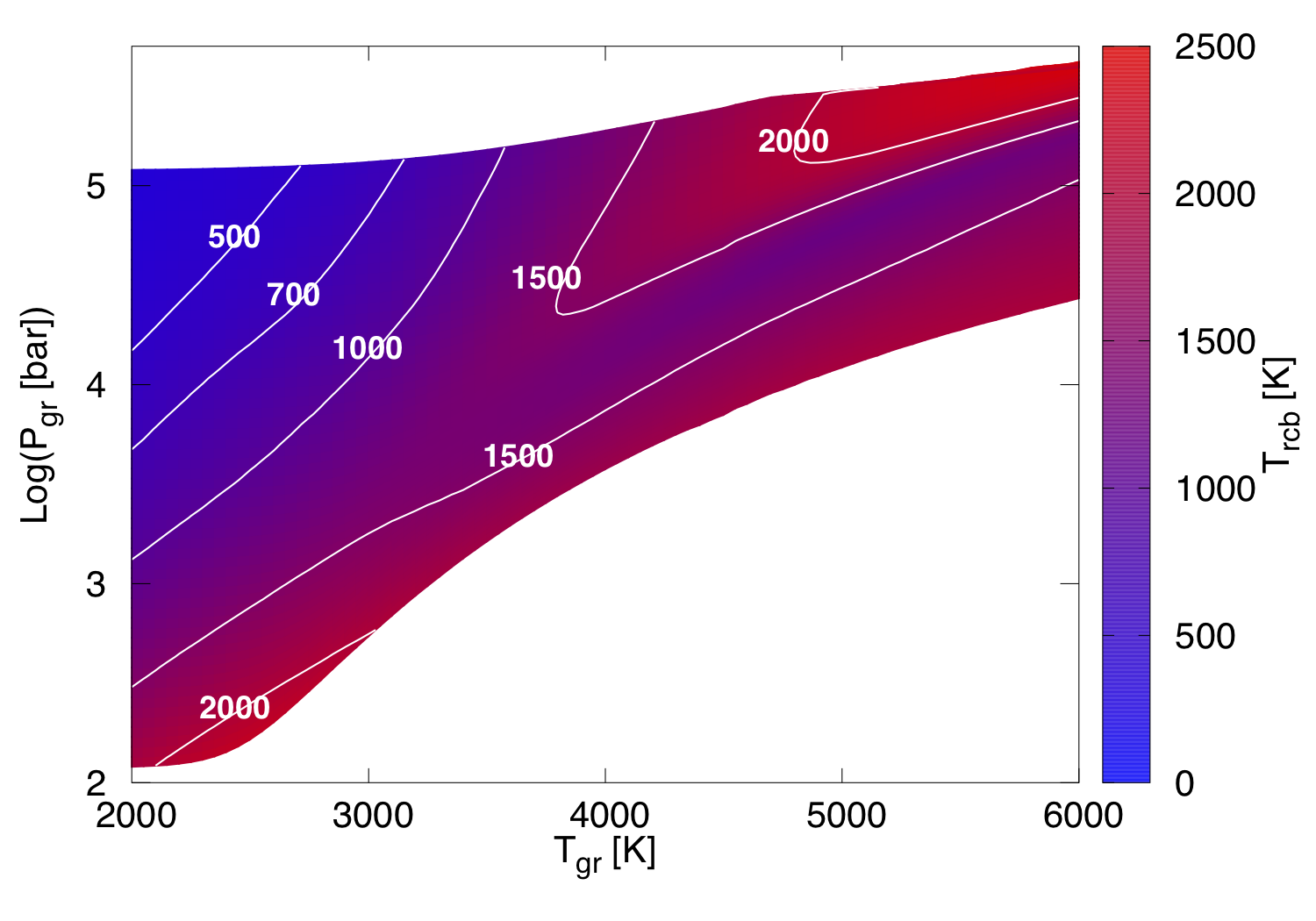}
   \end{center}
 \end{minipage}
 \begin{minipage}{1.0\textwidth}
    \begin{center}   
   ($b$) $P_\mathrm{rcb}=10$~bar \\
\includegraphics[width=0.495\textwidth]{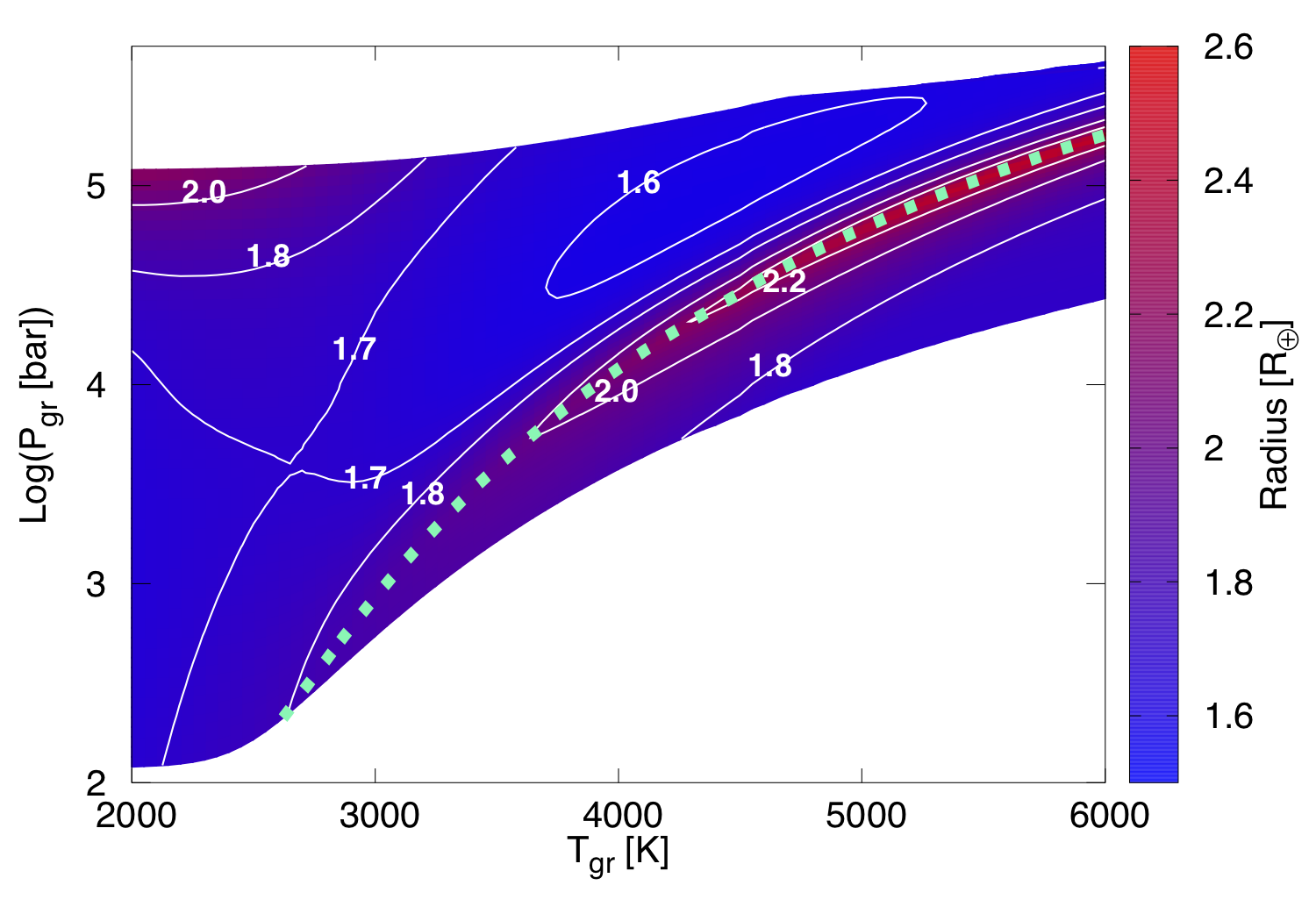}
\includegraphics[width=0.495\textwidth]{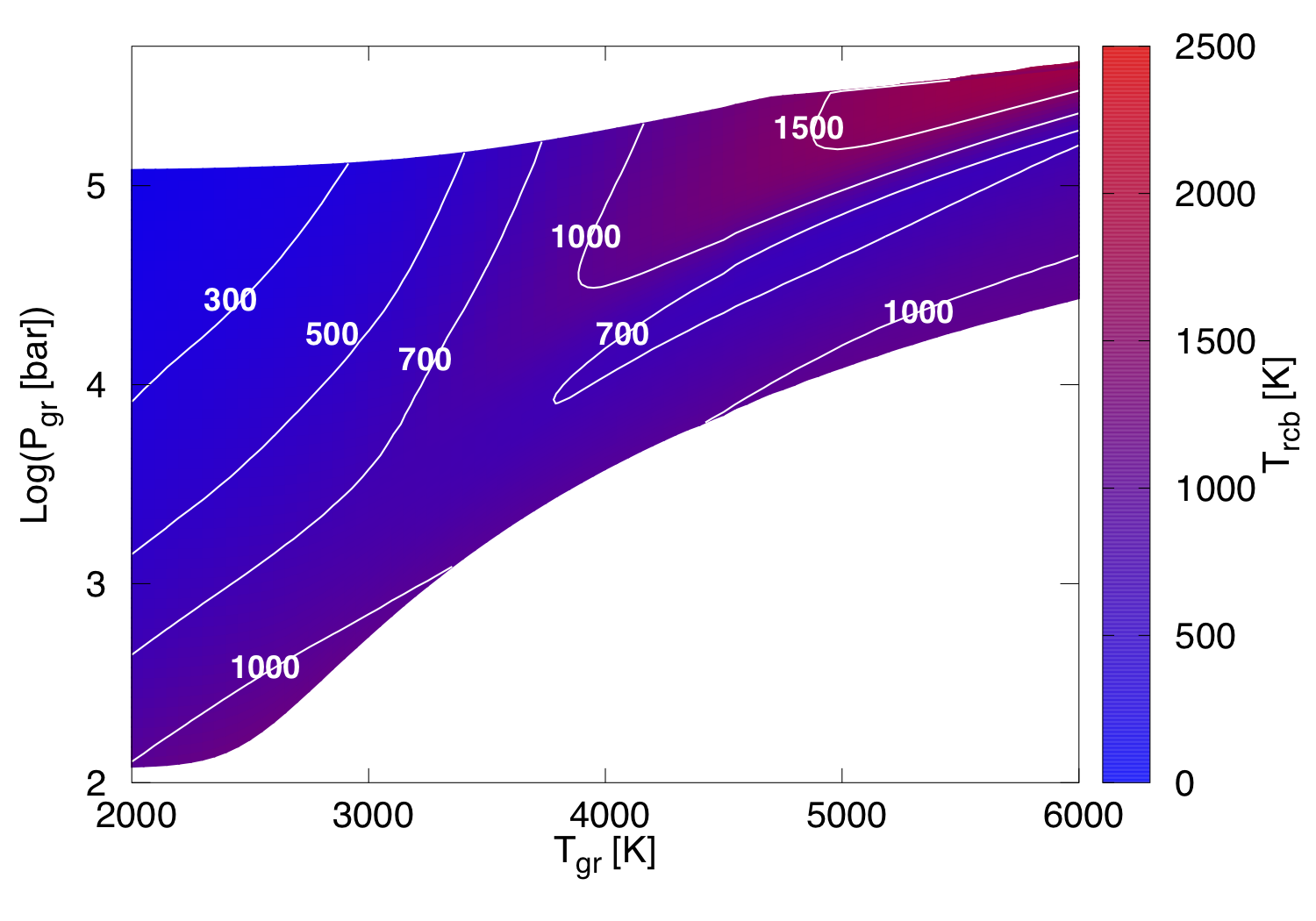}
   \end{center}
 \end{minipage}
  \begin{minipage}{1.0\textwidth}
    \begin{center}   
   ($c$) $P_\mathrm{rcb}=1$~bar \\
\includegraphics[width=0.495\textwidth]{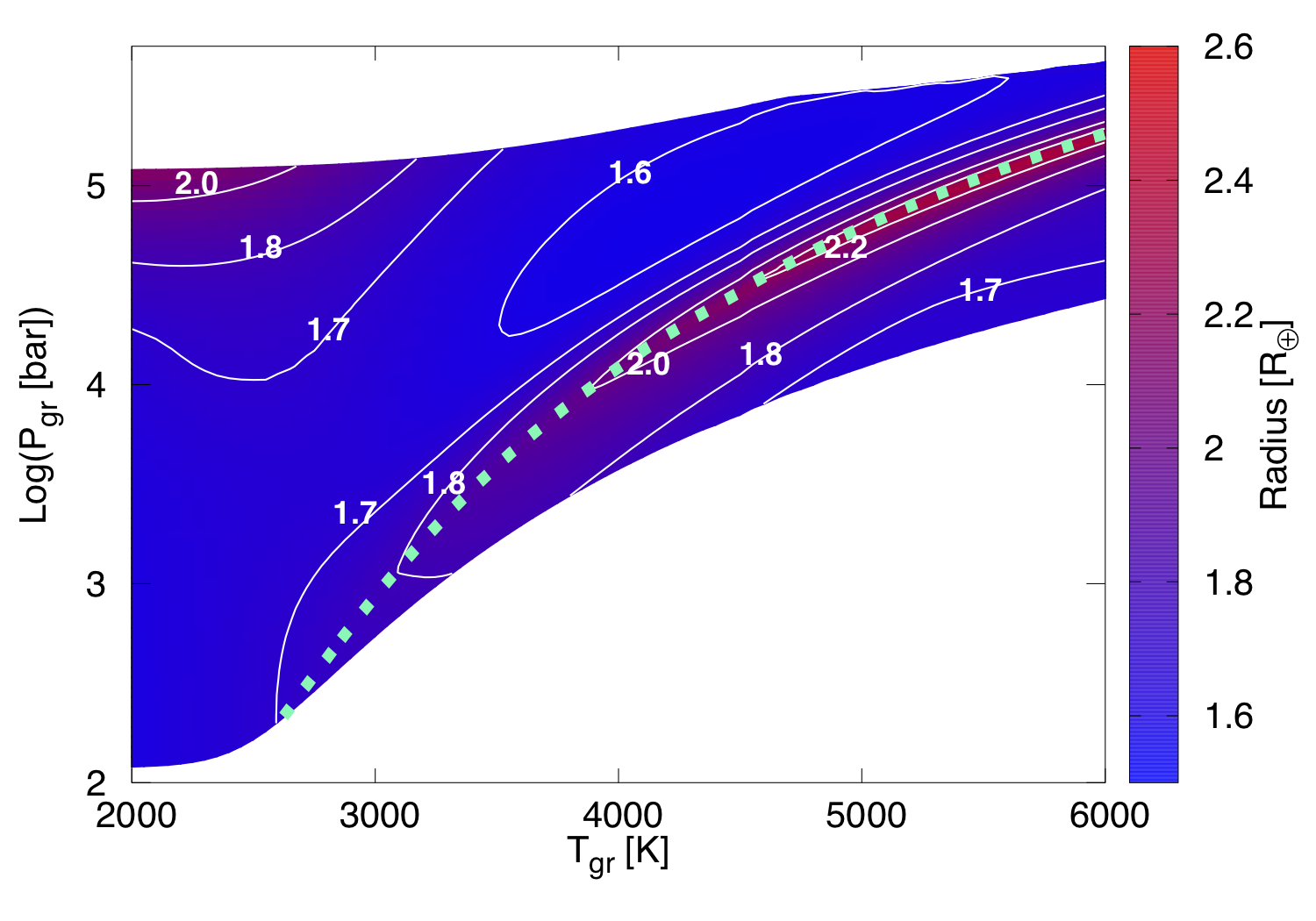}
\includegraphics[width=0.495\textwidth]{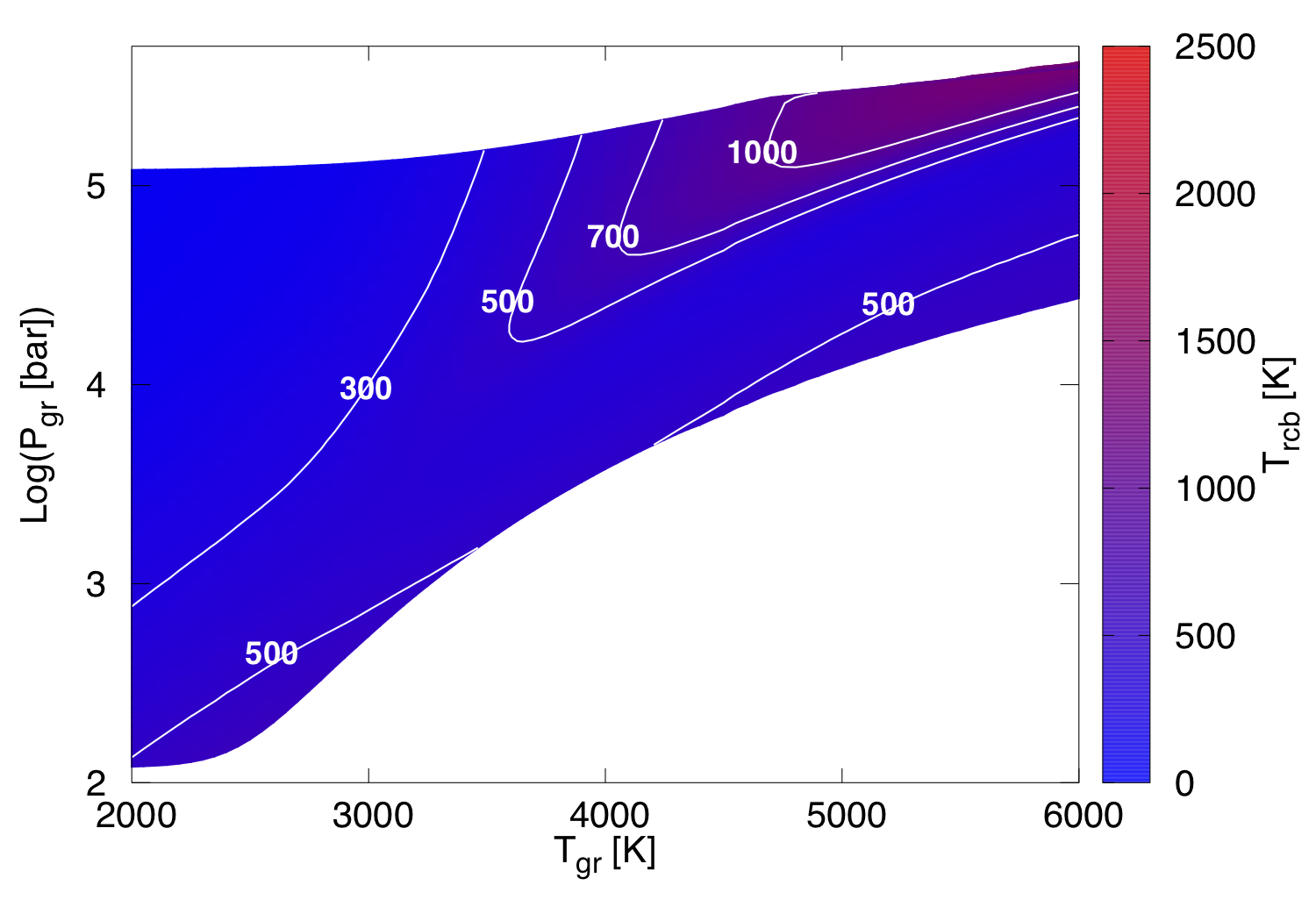}
   \end{center}
 \end{minipage}

\caption{ 
Planetary radius at 0.1 bar (left column) and tropopause temperature (right column)  for different ground pressure, $P_\mathrm{gr}$, and temperature, $T_\mathrm{gr}$. For the pressure of radiative convective boundary, $P_\mathrm{rcb}$, we assume 100~bar ($a$), 10~bar ($b$) and, 1~bar ($c$).
Dotted lines in left column represent the condition with a Si/O ratio of 1 at the ground.
White regions in all panels  represent areas outside the calculated parameter space.}
\label{fig:atom_compy}
\end{figure*}

Here, we show the atmospheric composition under different conditions of magma ocean surfaces, parameterized by the ground temperature, $T_\mathrm{gr}$, and the ground hydrogen pressure, $P_\mathrm{H_2, gr}$. Figure~\ref{fig:atom_compx} shows the molar fractions of SiH$_4$ at 0.1~bar ($a$) and at the ground ($b$), and that of H$_2$O ($c$) and SiO ($d$) at the ground, as well as the H$_2$O mass concentration in the magma ocean ($e$), under various ground pressures and ground temperatures for $P_\mathrm{rcb}=10$~bar. Also, Figure~\ref{fig:atom_compx}$f$ shows the relationship between the ground pressure of hydrogen, $P_\mathrm{H_2, gr}$, which is a given parameter in our model, and the resultant ground pressure.
{
Before discussing the abundance of SiH$_4$ at 0.1~bar (Fig.~\ref{fig:atom_compx}$a$), we first describe the behavior of species at the ground.}

{
The ground hydrogen pressure is a key parameter increasing the total ground pressure and controlling atmospheric composition. In particular, it is the dominant contributor to the total pressure (i.e., $P_\mathrm{H_2,gr}\sim P_\mathrm{gr}$) for $T_\mathrm{gr}\lesssim 2500$~K while its contribution decreases at higher temperature, as shown in Fig.~\ref{fig:atom_compx}$f$. 
For example, when $P_\mathrm{H_2,gr} = 10^2$, $10^3$, or $10^4$bar, the total ground pressure $P_\mathrm{gr}$ remains constant at $\chi_\mathrm{H_2}^{-1} P_\mathrm{H_2,gr}$ (see Eq.\ref{eq:Mh2}) for $T_\mathrm{gr} \lesssim 2500$, 3000, and 3500~K, respectively, but increases with $T_\mathrm{gr}$ above these temperatures.
This increase in $P_\mathrm{gr}$ with $T_\mathrm{gr}$ at a fixed $P_\mathrm{H_2,gr}$ is caused by the evaporation of SiO$_2$ (R1) which produces more SiO at higher temperature.
As shown in Fig.~\ref{fig:atom_compx}$d$,
the molar fraction of SiO at the ground increases with $T_\mathrm{gr}$, and SiO becomes the most abundant species in regions where its fraction exceeds $10^{-0.3}$ ($\sim50$~\%). 
Note that the abundant SiO does not remain in the upper atmospheric layers in our model due to the condensation of silicates.
}

{
The atmospheric H$_2$O fraction at the ground and the dissolved H$_2$O concentration in the magma ocean 
 increase with $T_\mathrm{gr}$ and $P_\mathrm{gr}$, as shown in Fig.~\ref{fig:atom_compx}$c$ and $e$. Both quantities exhibit a similar increasing trend, since the dissolved H$_2$O concentration increases with the atmospheric H$_2$O pressure following the solubility law (Eq. \ref{eq:diss}).
Their increase with $T_\mathrm{gr}$ results from the evaporation of SiO$_2$ (R1) which produces more O$_2$ at higher temperatures and thereby promotes the formation of H$_2$O (R2). The dependence on $P_\mathrm{gr}$ primarily reflects the increase in $P_\mathrm{H_2,gr}$, since  the formation of H$_2$O (R2) is also enhanced by higher $P_\mathrm{H_2,gr}$.  Due to the combined effects of these two factors, in the case of the atmospheric H$_2$O fraction
 at $P_\mathrm{gr} = 10^4$~bar, the H$_2$O fraction increases from the 0.01\%-level to the 0.1\%-level with $T_\mathrm{gr}$ for $T_\mathrm{gr} \lesssim 3000$~K, 
 but slightly decreases at even higher temperature, as the hydrogen pressure decreases with increasing temperature at $P_\mathrm{gr} = 10^4$~bar (see Fig.~\ref{fig:atom_compx}$f$).
}

{
The molar fraction of SiH$_4$ at the ground also depends on two factors: $T_\mathrm{gr}$ and $P_\mathrm{H_2,gr}$. In Fig.~\ref{fig:atom_compx}$b$, it increases with $T_\mathrm{gr}$, while the decline in the SiH$_4$ fraction is observed when $P_\mathrm{gr} > P_\mathrm{H_2,gr}$ (see also Fig.~\ref{fig:atom_compx}$f$). This dual dependence naturally arises from the reaction producing SiH$_4$ from SiO and H$_2$ (R4).
The former dependence primarily originates from the evaporation of SiO$_2$ (R1), which produces more SiO at higher temperature, whereas the latter is mainly due to the decrease in the H$_2$ pressure in this region.
In particular, 
the high sensitivity of the SiH$_4$ molar fraction to $P_\mathrm{H_2,gr}$ can be explained by the relation,
$x'_\mathrm{SiH_4}\propto P_\mathrm{H_2,gr}^4 P_\mathrm{gr}^{-3} {x'_\mathrm{H_2O}}^{-2}$,
which is derived from Eqs.~(\ref{eq:x_sih4}) and (\ref{eq:x'x}). Although the derived relation includes a dependence on the H$_2$O fraction, the variation in the H$_2$O molar fraction is minor compared to the change in $P_\mathrm{H_2,gr}$ around the transition region in the SiH$_4$ fraction, as shown in Fig.~\ref{fig:atom_compx}$c$ and $f$, respectively.
This indicates that the decline in the SiH$_4$ fraction is primarily driven by the reduction in  $P_\mathrm{H_2,gr}$.
}

{
Based on the behavior of species at the ground explained above, we now describe the molar fraction of SiH$_4$ at 0.1~bar, where it becomes observationally accessible. 
In many cases within the parameter space of $P_\mathrm{H_2, gr}$ and $T_\mathrm{gr}$ that we explored, SiH$_4$ is present at 0.1 bar with a molar fraction exceeding 0.1~\%. Figure~\ref{fig:atom_compx}$a$ illustrates an increase in the molar fraction of SiH$_4$ at 0.1~bar with the ground temperature,
 following the same trend as that at the ground, which originates from the evaporation of SiO$_2$ (R1).
}

{
Regarding the conditions under which SiH$_4$ is present in the upper atmospheric layers, pressure boundaries associated with a sharp decline in the SiH$_4$ abundance at 0.1 bar are observed, where the conditions at its molar fraction of 10$^{-3}$ and lower are nearly identical in Fig.~\ref{fig:atom_compx}$a$.
 The boundaries are caused by the condensation of SiO(c) (R5), which dominates over that of SiO$_2$(c) (R4) in our calculations. As a result, the pressure boundaries align with a Si/O ratio of
1 at the ground (dotted line in Fig~\ref{fig:atom_compx}$a$). A Si/O ratio of unity corresponds to the condition where the molar fractions of SiH$_4$ and H$_2$O become equal at the ground.
Accordingly, the pressure boundaries fall within the region where the SiH$_4$ fraction at the ground decreases with decreasing $P_\mathrm{H_2,gr}$ (Fig.~\ref{fig:atom_compx}$b$).  Above the pressure boundaries for the presence of SiH$_4$ at 0.1~bar, the molar fraction of SiH$_4$ increases with $P_\mathrm{H_2, gr}$, exceeding 10~\% for $T_\mathrm{gr}\gtrsim3200$~K.}

{
The presence of SiH$_4$ at the  upper atmospheric layers shown in Fig.\ref{fig:atom_compx}$a$ is 
 a consequence of  incorporating water dissolution in our model, which is also demonstrated in Fig.~\ref{fig:atom_comp1} and Fig.~\ref{fig:atom_comp2}.
We discuss the effect on  the overall H-O-Si chemical system considered in our model, in comparison with \citet{Misener+2023}, in Section~\ref{sec:d_wde}.  
Also, our adopted solubility law parameter values would leads to an underestimation of the SiH$_4$ molar fraction, compared to that calculated with the other values summarized in \citet{Bower+2022}, as discussed in Section~\ref{sec:d_sol}.
}

In addition, we show how the planetary radius and the temperature at the upper atmospheric layer,  which are observable quantities, are influenced by the abundant SiH$_4$ in the atmospheres. 
Figure~\ref{fig:atom_compy} shows the optical planetary radii and tropopause temperatures of atmospheres with tropopause pressures of 100~bar ($a$), 10~bar ($b$) and 1~bar ($c$)
under various ground pressures and ground temperatures. 
Regardless of the effect of SiH$_4$, planetary radii generally increase with ground pressure while  $T_\mathrm{rcb}$ decreases with ground pressure for identical atmospheric composition and ground temperature. This trend arises because higher ground pressure leads to a more extended atmosphere, increasing the radius and the temperature difference between the ground and the tropopause, which in turn lowers $T_\mathrm{rcb}$.
For example, 
under conditions where the
 SiH$_4$ fraction at 0.1~bar is 10\%  at $T_\mathrm{gr}\sim$3200~K (see  Fig.~\ref{fig:atom_compx}$a$)
 {, our calculation for $P_\mathrm{rcb}=100$~bar shows that the radius increases with $P_\mathrm{gr}$ from 1.7~$\Rearth$ to 1.8~$\Rearth$, while $T_\mathrm{rcb}$ decreases with $P_\mathrm{gr}$, falling below 1000~K (see Fig.~\ref{fig:atom_compy}$a$).}

Beyond the trends, the presence of abundant SiH$_4$ introduces additional effects. Since SiH$_4$ increases the mean molecular weight of the atmosphere, as shown in Fig~\ref{fig:atom_comp3}, it reduces the atmospheric scale height, leading to a smaller planetary radius. In addition, SiH$_4$ has a higher heat capacity than H$_2$ and He \citep[see][]{Chase1998}, which contributes to a higher $T_\mathrm{rcb}$. These effects are particularly evident in cases where the SiH$_4$ fraction exceeds 10~\% at 0.1~ bar. Under such conditions {, even with  $P_\mathrm{gr}$ higher than 10$^4$~bar and $T_\mathrm{gr}$ above 3200~K which would generally lead to a larger radius than that at $P_\mathrm{gr} < 10^4$~bar and $T_\mathrm{gr} < 3200$~K if SiH$_4$ were absent}, our calculations for $P_\mathrm{rcb}=$100~bar indicate a minimum planetary radius of less than 1.7~$R_\oplus$ and a maximum $T_\mathrm{rcb}$ exceeding 2000 K (Figure~\ref{fig:atom_compy}$a$). This behavior results from the significant fraction of SiH$_4$ throughout the atmosphere.
 {In particular, this suggests that the influence of the abundant SiH$_4$ on the planetary radius is greater than that of $T_\mathrm{gr}$ and $P_\mathrm{gr}$ alone.}
We note that another maximum value in $T_\mathrm{rcb}$ higher than 2000~K for $P_\mathrm{gr}<10^3$~bar in  Figure~\ref{fig:atom_compy}$a$ is primarily caused by the low atmospheric pressure rather than by SiH$_4$.

Another parameter in our model, $P_\mathrm{rcb}$, affects the planetary radius and $T_\mathrm{rcb}$, although it does not change the abundance of SiH$_4$ for given $P_\mathrm{H_2, gr}$ and $T_\mathrm{gr}$. This is because $P_\mathrm{rcb}$ changes only the temperature-pressure profile of the upper atmospheric layer with the given surface condition.
Thus, the smaller $P_\mathrm{rcb}$ leads to a smaller planetary radius and a smaller $T_\mathrm{rcb}$, as shown in Figure~\ref{fig:atom_compy}. 
{In this parametrization, the planetary radius is relatively insensitive to $P_\mathrm{rcb}$, whereas $T_\mathrm{rcb}$ shows a strong dependence on it. The maximum and minimum radii for the SiH$_4$-rich atmospheres
differ only slightly, ranging from about 2.2-2.4~$R_\oplus$ to 1.6-1.7~$R_\oplus$ for $P_\mathrm{rcb}=$1--100~bar, as shown in the left panels of Figure~\ref{fig:atom_compy}.
 In contrast, the maximum value of $T_\mathrm{rcb}$ vary more significantly,  from 1000~K to 2000~K for $P_\mathrm{rcb}=$1--100~bar, as shown in the right panels of Figure~\ref{fig:atom_compy}.
} 
Since the radius and temperature of the upper atmospheric layer are observable, they are crucial for identifying exoplanets that 
may have SiH$_4$-rich atmospheres. 
However, our model may oversimplify the thermal structure. 
This limitation is further discussed in Sections~\ref{sec:d_tp} and \ref{sec:sns}.
Moreover, our model does not account for the non-ideal behavior of SiH$_4$, which could lead to an overestimate of the planetary radius, as discussed in Section~\ref{sec:d_non}.

\section{Discussion} \label{sec:disc}

\subsection{Redox state of magma ocean}
\label{sec:d_r}

\begin{figure}[t]
 \begin{minipage}{0.5\textwidth}
    \begin{center}
    {($a$) Redox state of magma ocean} \\
\includegraphics[width=\textwidth]{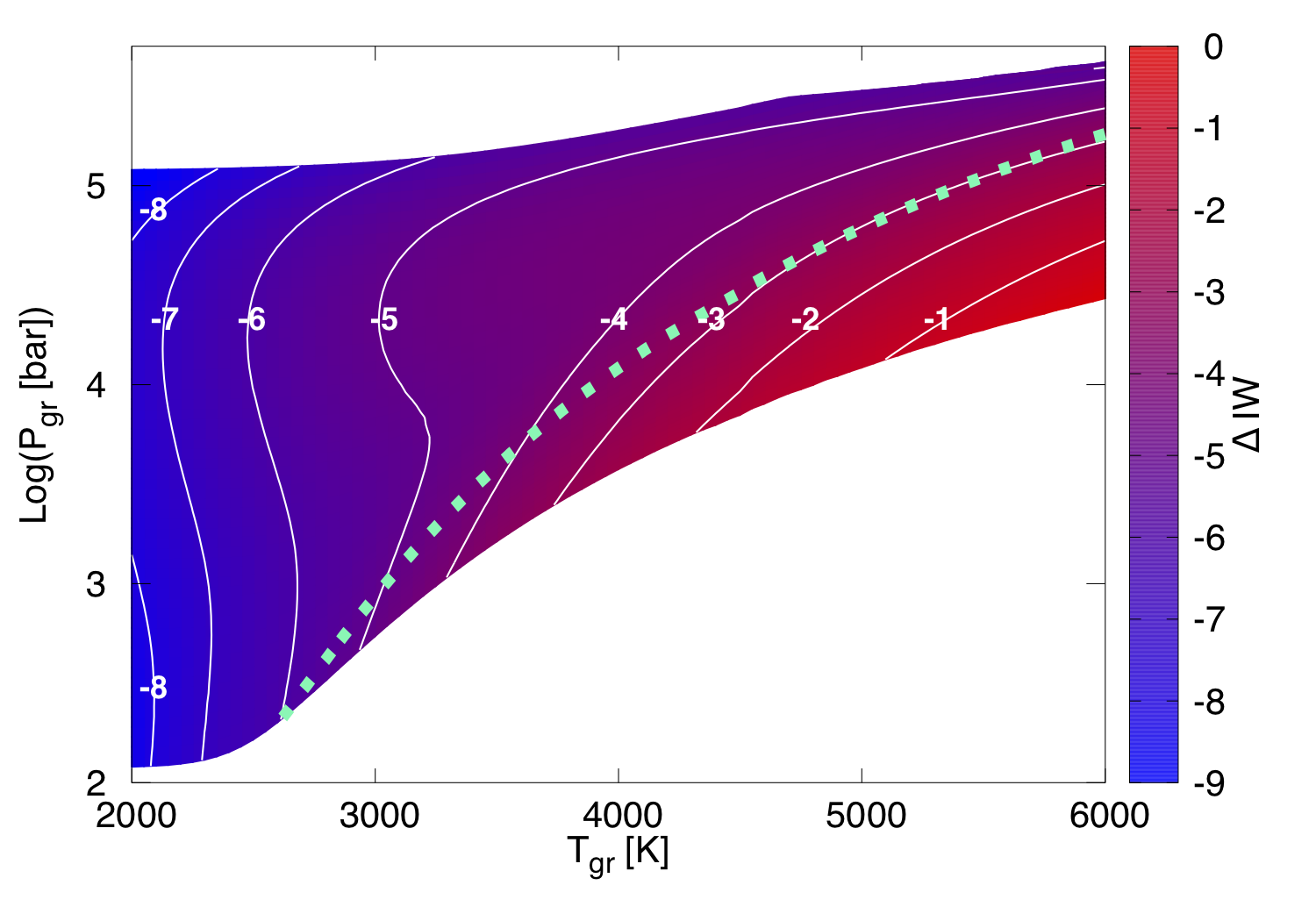}
   \end{center}
 \end{minipage}
  \begin{minipage}{0.5\textwidth}
    \begin{center}
     {($b$) O$_2$ ground pressure} \\
\includegraphics[width=\textwidth]{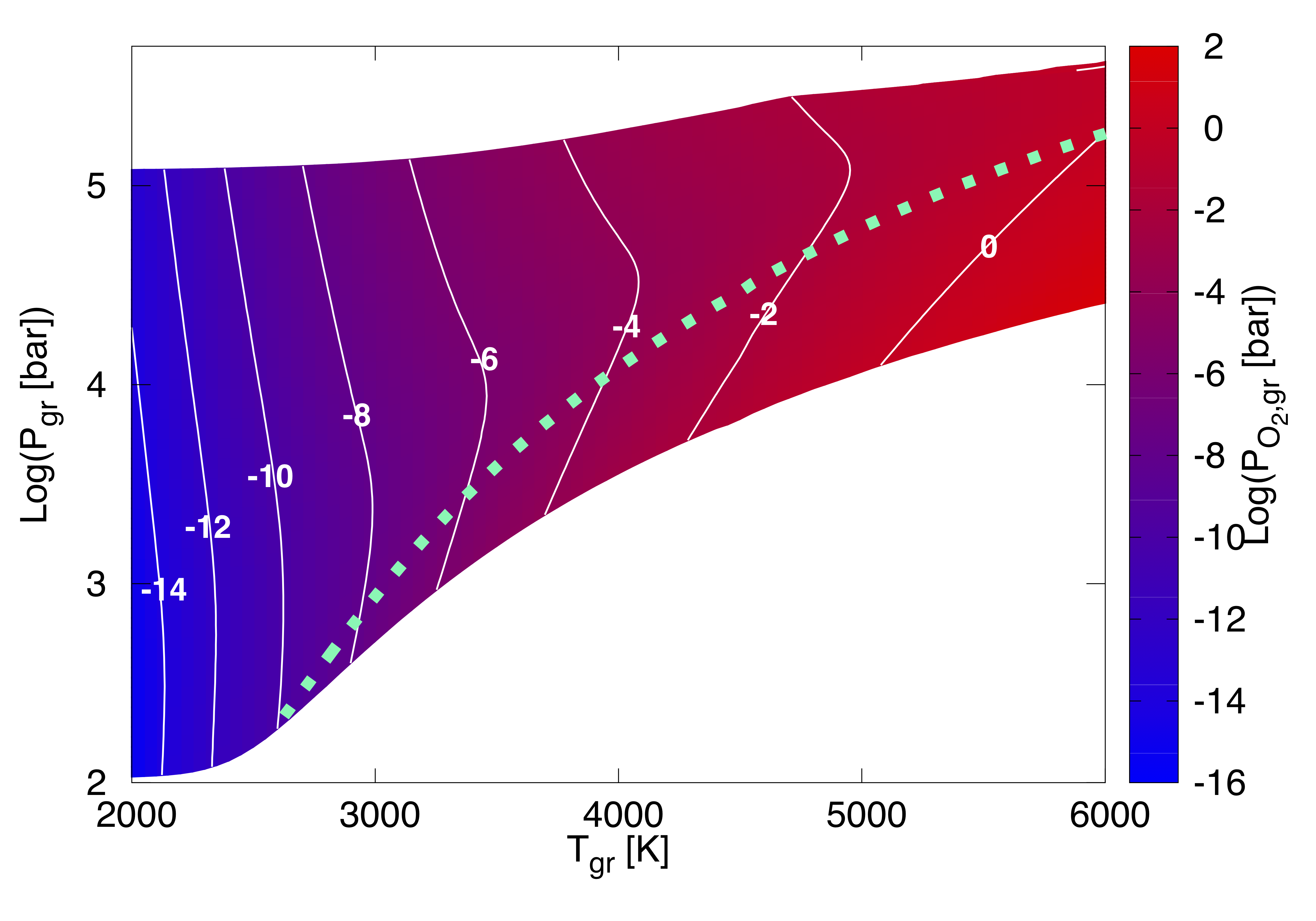}
   \end{center}
 \end{minipage}
\caption{Redox state of magma ocean {expressed as $\Delta$IW ($a$) and O$_2$ ground pressure ($b$)} for different ground pressure, $P_\mathrm{surf}$, and temperature, $T_\mathrm{gr}$. Color counter and lines show the value of $\Delta$IW calculated from 
{the}
oxygen pressure 
and the reported function of the oxygen fugacity of iron--w\"{u}stite buffer \citep{Hirschmann+2008} {($a$), and the pressure in bar ($b$)}.
Dotted line represents a boundary of SiH$_4$ presence and absence at 0.1 bar in Fig.~\ref{fig:atom_compx}$a$.}
\label{fig:atom_compiw}
\end{figure}
Only oxygen molecules originating from the evaporation of SiO$_2$ magma ocean have been considered in our model.
This corresponds to highly reduced magma composition compared to Earth-like rock.
In Earth-like rock, Fe-oxides such as FeO and Fe$_2$O$_3$ are thought to play a central role
controlling the oxygen amount \citep[e.g.,][]{Frost+2008}.
A recent experiment indicated that a magma ocean with a bulk silicate Earth-like composition provided an oxygen fugacity of 0.5 log units above 
the iron--w\"{u}stite (IW) buffer ($\Delta$IW=+0.5)
 \citep{Sossi+2020}.
If a sub-Neptune has an Earth-like oxidized magma ocean, abundant H$_2$O would be produced through the reaction between atmospheric H$_2$ and the magma ocean, as demonstrated in previous studies \citep[e.g.,][]{Kite+2020,Tian+2024,Seo+2024}. In such water-rich atmospheres, SiH$_4$ is expected to be depleted via the reaction R3, followed by the condensation of silicates, as shown in Fig.~\ref{fig:atom_comp1}$b$.

Figure~\ref{fig:atom_compiw}{$a$} illustrates the redox state of magma ocean in our model, showing the value of $\Delta$IW calculated from our resultant oxygen pressure at the ground. {For reference, the oxygen pressure itself is also shown in Fig.~\ref{fig:atom_compiw}$b$}. For $\Delta$IW, we 
use the temperature-pressure-dependent function of oxygen fugacity for the iron--w\"{u}stite buffer which is inferred from experiments by \citet{ Hirschmann+2008} whilst assuming the fugacity coefficient of oxygen to be unity for our model consistency.  
It should be noted that we extrapolate the function of the iron--w\"{u}stite's oxygen fugacity to higher temperature than 3000~K, which exceeds the experimentally investigated range \citep{Hirschmann+2008}.
From Fig.~\ref{fig:atom_compiw}{$a$} and \ref{fig:atom_compx}$a$, a reduced magma ocean with a $\Delta$IW value of less than -3 is necessary to sustain abundant SiH$_4$ throughout the atmosphere. 
Such highly reduced composition would be feasible if FeO-free magma exists. For example, experimental studies reported a very low oxygen fugacity of $\sim10^{-20}$~bar at $\sim$1400K for Enstatite condrite 
\citep[][]{Fogel+1989}, which corresponds to $\Delta$IW$\sim-6$. 

The redox state of a magma ocean is expected to be determined by the composition of its building blocks during planetary formation and the differentiation of the  innermost iron core \citep[e.g.,][]{Ringwood+1977,Wanke+1981,Kuramoto+1996}. While Earth's current rocks are oxidized, the isotopic composition of Earth suggests that Earth was built primarily from enstatite chondrites  \citep{JAVOY+2010,WARREN2011,Dauphas+2017}.
However, even if planets initially formed with such reduced materials, the redox state of the magma ocean could be altered by core differentiation,
which depends on factors such as iron droplet size, heat flow, and melt fraction \citep{Elkins-Tanton+2008}. 

If differentiation is inefficient, as suggested for sub-Neptunes by \citet{Lichtenberg+2021}, iron droplets may remain suspended in a turbulent magma ocean, delaying core formation. In such a scenario, equilibration between dissolved water in hydrous silicate melt, which is considered in this study, and iron droplets could lead to  the formation of FeO \citep{Elkins-Tanton+2008}.
Furthermore, under very high pressure and temperature, silicates can decompose into neutral Si, which dissolves into the iron core as a metal \citep{FISCHER+2015}, also leading to the oxidation of the magma ocean. Conversely, an FeO-free reduced magma ocean might be produced if planets formed with reduced materials such as enstatite chondrites, the differentiation of the innermost iron core was efficient, and the iron core became isolated from the outer silicate portion. Although this scenario might be unlikely \citep{Lichtenberg+2021}, whether it could occur for sub-Neptunes or not remains uncertain. 

We note that our model, particularly the assumption of a SiO$_2$ magma ocean, may be invalid under conditions where the $\Delta$IW value is less than $-7$  (see Fig.~\ref{fig:atom_compiw}{$a$}). This is because the metallization of Si could occur under such highly reduced conditions, leading to oxygen fugacity being buffered by the Si-SiO$_2$ system at $\Delta$IW$\sim-7$ \citep{Seidler+2024}. In such cases, the equations of mass balance in Sec.~\ref{sssec:mb} would need to include the term of metallic Si, although investigating such states is beyond the scope of this study.

\subsection{Thickness of magma ocean}
\label{sec:d_t}

We have demonstrated the impact of water dissolution into magma on the atmospheric SiH$_4$ abundance, as shown in Figs.~\ref{fig:atom_comp1} and \ref{fig:atom_comp2}. 
The effect of water dissolution depends on not only the solubility law (Eq.~\ref{eq:diss}) but also the volume of magma ocean within the interior core.
 In all our results shown in Sec.~\ref{sec:res}, we assume deep magma ocean, adopting 0.5 as the magma fraction of the interior core, $\zeta$. To assess the sensitivity of our results to the magma ocean fraction, we explore cases with different values of $\zeta$.
 
\begin{figure}[t]
 \begin{minipage}{0.5\textwidth}
    \begin{center}
    ($a$) $\zeta=0.1$
\includegraphics[width=\textwidth]{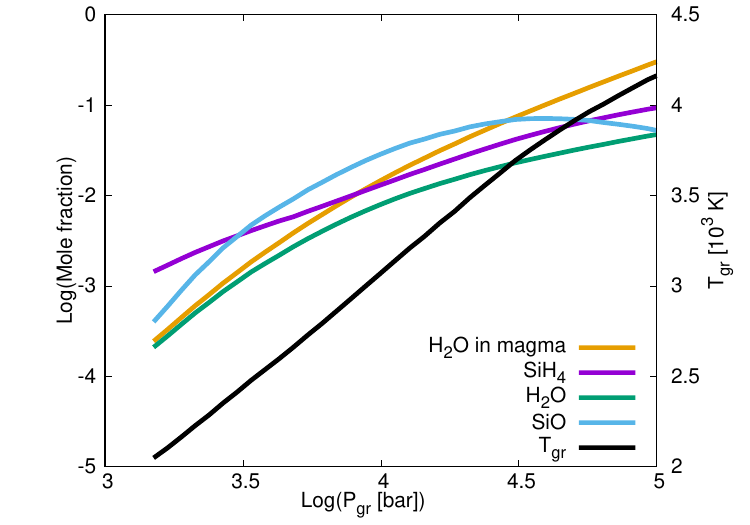}
   \end{center}
 \end{minipage}
  \begin{minipage}{0.5\textwidth}
    \begin{center}
    ($b$) $\zeta=0.01$
\includegraphics[width=\textwidth]{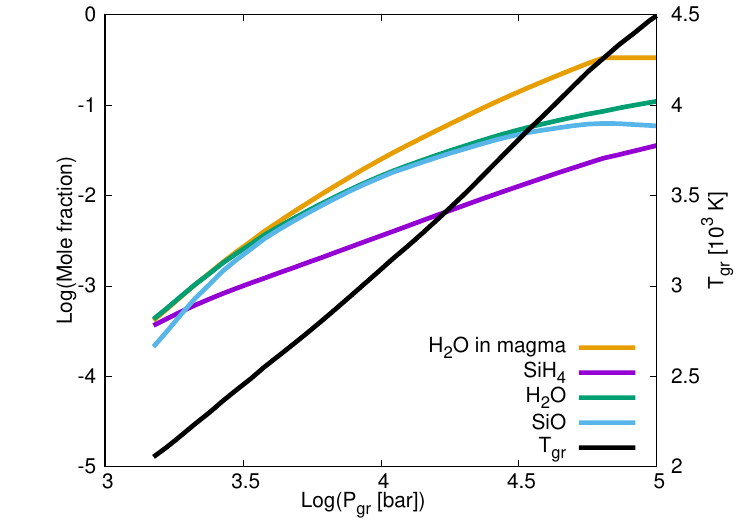}
   \end{center}
 \end{minipage}
\caption{Molar fractions of 
H$_2$O in the magma ocean ($=X_\mathrm{H_2O}\times 60/18$) (orange),  
atmospheric SiH$_4$ (purple), H$_2$O (green), and SiO (cyan) at the ground, and the ground temperature, $T_\mathrm{gr}$ (black),  are shown as functions of the ground pressure, $P_\mathrm{gr}$, for $T_\mathrm{rcb}=500$~K and $P_\mathrm{rcb}=10$~bar. Upper and lower panels show the results obtained for $\zeta=0.1$ ($a$) and 0.01 ($b$), respectively.}
\label{fig:atom_comp01}
\end{figure}

Figure~\ref{fig:atom_comp01} shows the atmospheric composition at the ground for $\zeta=0.1$ ($a$) and 0.01 ($b$) under various ground pressure with  $T_\mathrm{rcb}=500$~K and $P_\mathrm{rcb}=10$~bar. These conditions are the same as those used in our nominal calculation for $\zeta=0.5$, shown in Fig.~\ref{fig:atom_comp2}. Also, $\zeta = 0.1$ and $\zeta = 0.01$ correspond to magma oceans with thickness of approximately 3~\% and 0.3~\% of the core radius, respectively.
For $\zeta = 0.1$, 
the faction of SiH$_4$ is lower than that of SiO for $P_\mathrm{gr}$ from $10^{3.5}$~bar to 10$^{4.7}$~bar. 
However, it is larger than that of H$_2$O over all $P_\mathrm{gr}$ for $\zeta = 0.1$
 (Fig.~\ref{fig:atom_comp01}$a$). In contrast, for $\zeta=0.01$,  the faction of SiH$_4$ is lower than that of H$_2$O  at all $P_\mathrm{gr}$ (Fig.~\ref{fig:atom_comp01}$b$).
Therefore, in the case of $T_\mathrm{rcb}=500$~K and $P_\mathrm{rcb}=10$~bar, SiH$_4$ can persist in the upper atmospheres for  $\zeta\geq0.1$ but not for $\zeta=0.01$, considering the condensation of silicates.
The underlying reason for this trend is that a smaller magma ocean fraction corresponds to a smaller water reservoir size, which leads to a more oxidized atmosphere and, consequently, a decrease in the abundance of SiH$_4$ (see also Eq.~\ref{eq:mass_balance}). 
Additionally, based on Eq.~(\ref{eq:mass_balance}), our model assuming $\zeta=0.1$ or $\zeta=0.01$ provides the same results as a model in which 
 the solubility law of water is weakened by a factor of 5 or 50, respectively, for $\zeta=0.5$. 

\subsection{{Water dissolution}}
\label{sec:d_wd}
\subsubsection{{Effect on H-O-Si chemistry}}
\label{sec:d_wde}
{Here, we discuss the effect of water dissolution into the magma ocean on the overall H-O-Si chemical system considered in our model, although we primarily focused on its impact on SiH$_4$ in Sec~\ref{sec:res}.  
Under fixed $P_\mathrm{H_2, gr}$, water dissolution first leads to a lower $P_\mathrm{H_2O, gr}$ compared to the case without dissolution, which in turn results in a lower $P_\mathrm{O_2, gr}$ in the equilibrium of the H$_2$O formation reaction (R2).  
Subsequently, this decrease in $P_\mathrm{O_2, gr}$ promotes an increase in $P_\mathrm{SiO, gr}$, as explained by Le Chatelier’s principle for the evaporation of SiO$_2$ (R1).  
Finally, the combination of lower $P_\mathrm{H_2O, gr}$ and higher $P_\mathrm{SiO, gr}$ results in an enhanced $P_\mathrm{SiH_4, gr}$, following Le Chatelier’s principle for the formation reaction of SiH$_4$ (R4). Hence, water dissolution decreases the abundances of H$_2$O and O$_2$ but increases those of SiO and SiH$_4$ at the ground.
In particular, this chain of thermochemical responses highlights how water dissolution can indirectly favor SiH$_4$ formation through its influence on the redox state of magma ocean. 
}

{
As a result of this effect, our model predicts a lower H$_2$O fraction and higher SiO and SiH$_4$ fractions at the ground compared to those shown in  \citet{Misener+2023}, who did not account for water dissolution. 
\citet{Misener+2023} reports molar fractions of $\sim$20\% for H$_2$O, $\sim$8\% for SiO, and $\sim$5\% for SiH$_4$ at $P_\mathrm{H_2, gr} \sim 10^5$~bar and $T_\mathrm{gr} = 5000$~K (see their Fig.~4),  
whereas our model yields 2.4\%, 16\%, and 41\% for H$_2$O, SiO, and SiH$_4$, respectively, under the same conditions. 
Also, there is a slight difference in the total ground pressure due to the difference in composition, as  
our model yields $P_\mathrm{gr} \sim 3 \times 10^5$~bar, while the value inferred from \citet{Misener+2023} is approximately $1.5 \times 10^5$~bar\footnote{This inferred value is derived assuming $P_\mathrm{H_2, gr} = 10^5$~bar and an H$_2$ fraction of 67\%, based on the results presented in \citet{Misener+2023}.}.  
The differences in the SiO and SiH$_4$ fractions, resulting from the decreased H$_2$O abundance and the enhanced total pressure, can be explained by the following relationships:  
$x'_\mathrm{SiO} \propto P_\mathrm{H_2,gr} P_\mathrm{gr}^{-2} {x'_\mathrm{H_2O}}^{-1}$ and  
$x'_\mathrm{SiH_4} \propto P_\mathrm{H_2,gr}^4 P_\mathrm{gr}^{-3} {x'_\mathrm{H_2O}}^{-2}$,  
which are derived from Eqs.~(\ref{eq:x_sio}), (\ref{eq:x_sih4}), and (\ref{eq:x'x}). 
}

{In addition, 
the differences in these fractions lead to a difference in atmospheric mean molecular weight at the ground, which is $\sim10$~amu in \citet{Misener+2023} and $\sim20$~amu in our model. Such higher atmospheric mean molecular weight, combined with the presence of abundant SiH$_4$ throughout the atmosphere in our model,  could result in a smaller planetary radius. In the case of $P_\mathrm{H_2, gr} = 10^5$~bar and $T_\mathrm{gr} = 5000$~K, our model predicts a radius of 1.6--1.7$\Rearth$ for $P_\mathrm{rcb}=1$--100~bar (see Fig~\ref{fig:atom_compy}). Note that comparing the radius with that of \citet{Misener+2023} is not appropriate, as our model may oversimplify the atmospheric temperature profiles, as discussed in Section~\ref{sec:d_tp}.}

{
We note that the main conclusion of \citet{Misener+2023} and \citet{Misener+2022}, that enhanced silicon-bearing species due to interactions between the magma ocean and atmospheric hydrogen can inhibit atmospheric convection, would still hold even if water dissolution were incorporated into their model. 
This is because water dissolution enhances the SiO fraction at the ground, which could lead to a vertical gradient in the mean molecular weight through silicate condensation and this gradient could be sufficient to inhibit convection.  
Such convection inhibition may particularly occur in the parameter space where the ground SiO fraction exceeds 10\%, as shown in Fig.~\ref{fig:atom_compy}$d$, although exploring the detail is beyond the scope of this work. 
}

\subsubsection{{Sensitivity to solubility law}}
\label{sec:d_sol}
\begin{figure*}
 \begin{minipage}{0.5\textwidth}
    \begin{center}
    ($a$) $\alpha=2.15\times10^{-4}$~bar$^{-0.7}$ \& $\beta=0.7$
\includegraphics[width=\textwidth]{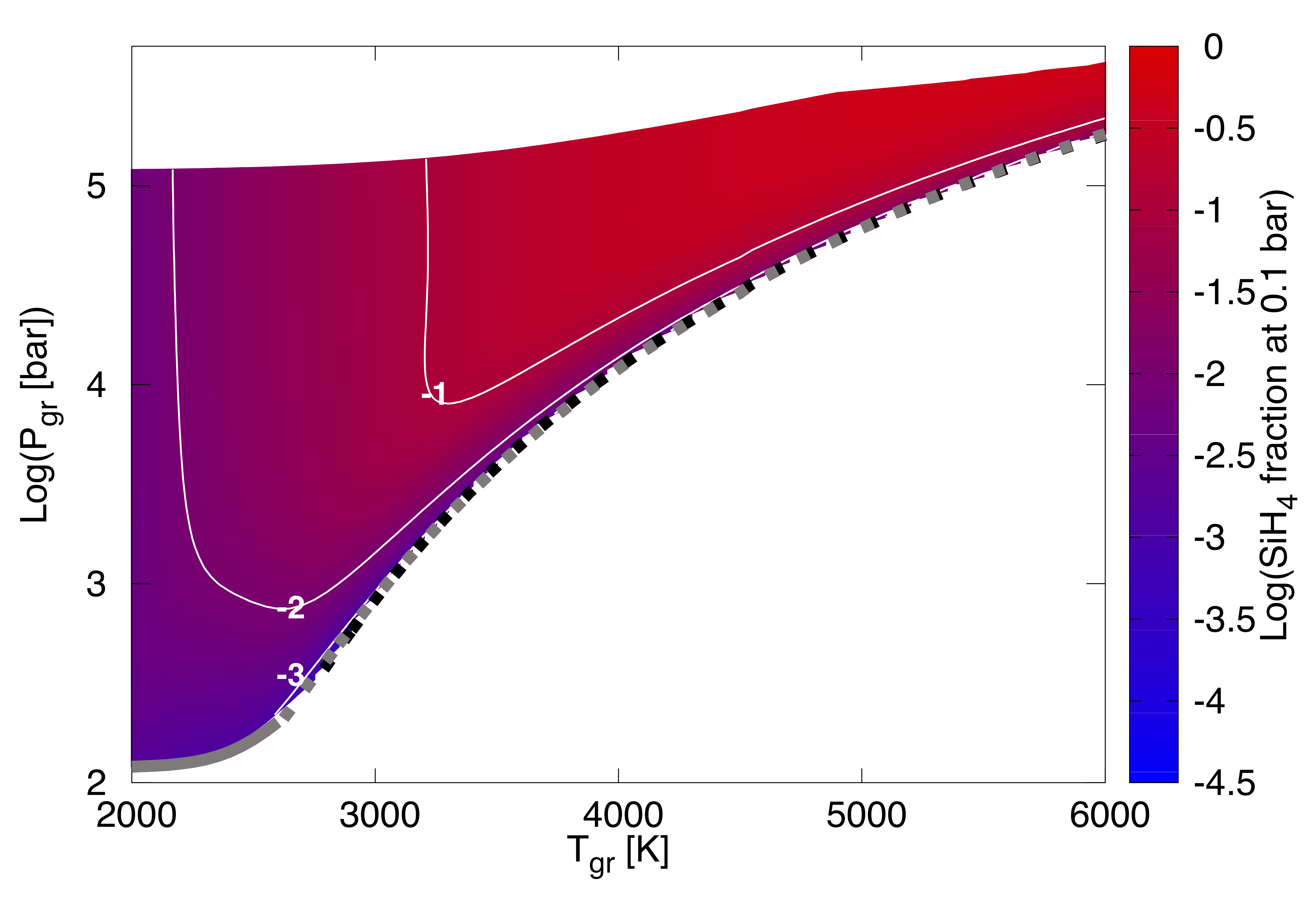}
   \end{center}
 \end{minipage}
 \\
  \begin{minipage}{0.5\textwidth}
    \begin{center}
    ($b$) $\alpha=5\times10^{-4}$~bar$^{-0.5}$ \& $\beta=0.5$
\includegraphics[width=\textwidth]{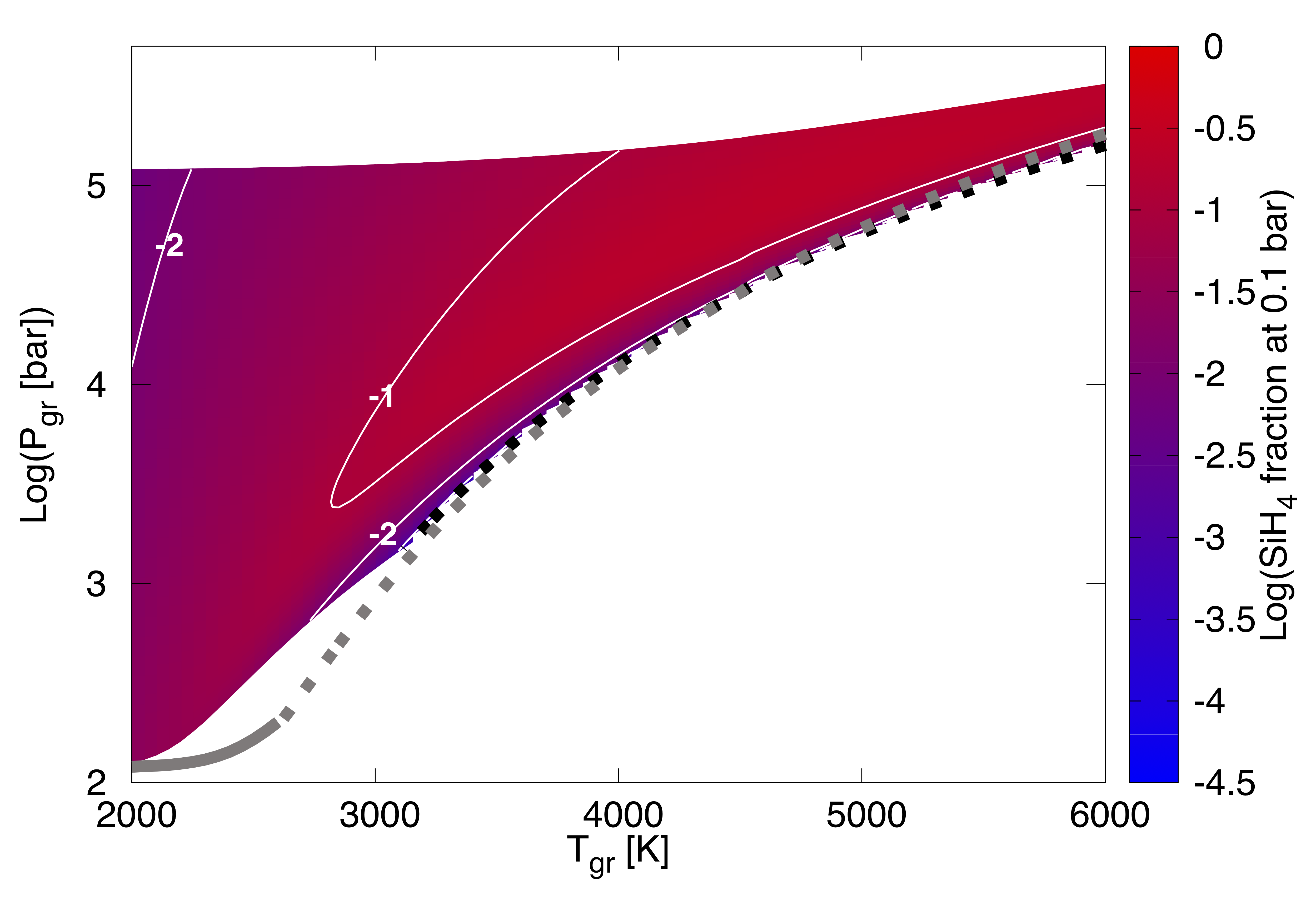}
   \end{center}
 \end{minipage}
   \begin{minipage}{0.5\textwidth}
    \begin{center}
    ($c$) $\alpha=10\times10^{-4}$~bar$^{-0.5}$ \& $\beta=0.5$
\includegraphics[width=\textwidth]{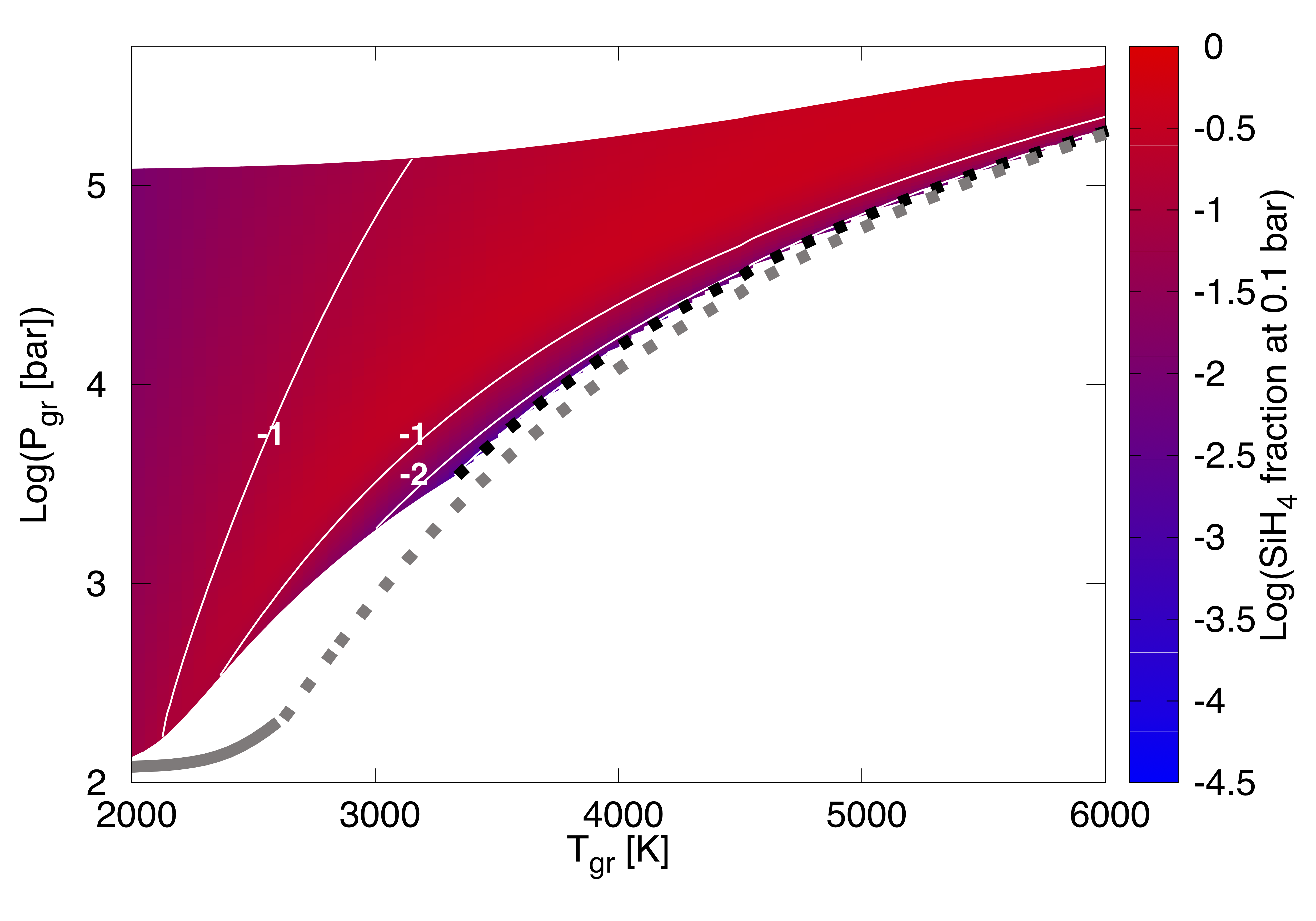}
   \end{center}
 \end{minipage}
\caption{{
Molar fractions of SiH$_4$ at 0.1~bar 
for different $\alpha$ and $\beta$ values in water solubility law: $2.15\times10^{-4}$~bar$^{-0.7}$ and $0.7$ ($a$), $5\times10^{-4}$~bar$^{-0.5}$ and $0.5$ ($b$), and  $10\times10^{-4}$~bar$^{-0.5}$ and $0.5$ ($c$).  Contour counters and lines show the log10 values of the fractions. Black dotted line in each panels represents the condition with a Si/O ratio of 1 at the ground. White regions represent areas outside the calculated parameter space and parameter spaces where no SiH$_4$ is present at 0.1~bar. For comparison, gray lines represents a boundary with $P_\mathrm{H_2, gr}=100$ (solid), and a boundary with a Si/O ratio of 1 at the ground (dotted) in Fig.~\ref{fig:atom_compx}$a$.}}
\label{fig:atom_comp_sol}
\end{figure*}

{
We have assumed $\alpha=3.44\times 10^{-4.3}$~bar$^{-0.74}$ and $\beta=0.74$ for the water solubility law (Eq.\ref{eq:diss}) in all our results presented in Sec.~\ref{sec:res}.
These values  have also been used in previous models for sub-Neptunes \citep{Kite+2020,Seo+2024}, and they yield  $X_\mathrm{H_2O}$ that agrees with experimental results for pure SiO$_2$ \citep{Kennedy1962,Holtz+2000} within 60~\%. 
In this section, we explore cases with different values of $\alpha$ and $\beta$ to assess the sensitivity of our results to the adopted values for water solubility parameters.
}

{
Among the solubility law parameters constrained by experimental studies summarized in a recent paper \citep{Bower+2022}, our adopted values are close to those derived for basalt at pressures of 1--6~kbar and a temperature of 1373~K by \citet{Willson+1981}, which give $\alpha = 2.15 \times 10^{-4}$~bar$^{-0.7}$ and $\beta = 0.7$. In contrast, other reported values are different from our adopted values, with $\beta = 0.5$ and $\alpha$ in the range of $5.34\times 10^{-4}$--$10.07 \times 10^{-4}$~bar$^{-0.5}$, derived for various compositions such as peridotite \citep{Sossi+2023}, lunar glass, anorthite-diopside mixtures \citep{Newcombe+2017}, and MORB-like basalt compositions \citep{Dixon+1995, Berndt+2002}, under pressures of 1~bar to 2~kbar and temperatures of 1473~K to 2173~K \citep[see Table 1 of][for the details]{Bower+2022}.}

{
In the range of H$_2$O partial pressure obtained in our results (Fig.~\ref{fig:atom_compx}$c$), which spans from $10^{-2}$~bar to $10^4$~bar, the $X_\mathrm{H_2O}$ values calculated using $\alpha = 2.15 \times 10^{-4}$~bar$^{-0.7}$ and $\beta = 0.7$ agree with those from our adopted values within 40\%. On the other hand, $X_\mathrm{H_2O}$ estimated using $\alpha = 5\times 10^{-4}$--$10 \times 10^{-4}$~bar$^{-0.5}$ and $\beta = 0.5$ is systematically higher than that from our adopted values at $P_\mathrm{H_2O} \lesssim 10^2$--$10^3$~bar, with the discrepancy increasing as the pressure decreases and reaching about one order of magnitude at $P_\mathrm{H_2O} = 10^{-2}$~bar.}

{
Therefore, in our model, the use of $\alpha$ and $\beta$ values adopted from \citet{Schaefer+2016} may lead to an underestimation of water solubility at the low pressures ($P_\mathrm{H_2O} \lesssim 10^2$--$10^3$~bar), even though they remain consistent with experiments conducted at $P_\mathrm{H_2O} = 1$--9~kbar \citep{Kennedy1962, Holtz+2000}. This underestimation of water solubility results in a smaller water reservoir size and thus a lower estimate for the SiH$_4$ fraction, as discussed in Sec.~\ref{sec:d_t}. In addition, it leads to a lower value of $P_\mathrm{gr}$ for a given $P_\mathrm{H_2, gr}$ when SiO is the dominant species contributing to $P_\mathrm{gr}$, as seen especially in the case of $P_\mathrm{H_2, gr} = 100$~bar and $T_\mathrm{gr}\gtrsim 2500$~K in Figs.~\ref{fig:atom_compx}$d$ and $f$. This occurs because the underestimated water solubility leads to a higher $P_\mathrm{H_2O, gr}$, which results in a lower $P_\mathrm{SiO, gr}$, as explained by Le Chatelier’s principle for reactions R1 and R2.
}

{
Figure~\ref{fig:atom_comp_sol} shows the molar fraction of SiH$_4$ at 0.1~bar for different water solubility parameter values, which are $\alpha=2.15\times10^{-4}$~bar$^{-0.7}$ and $\beta=0.7$ ($a$), $\alpha=5\times10^{-4}$~bar$^{-0.5}$ and $\beta=0.5$ ($b$), and  $\alpha=10\times10^{-4}$~bar$^{-0.5}$ and $\beta=0.5$ ($c$),
under various ground pressures and ground temperatures for $P_\mathrm{rcb}=10$~bar.  For $\alpha=2.15\times10^{-4}$~bar$^{-0.7}$ and $\beta=0.7$, the fraction of SiH$_4$ is  consistent with our nominal calculations (Fig.~\ref{fig:atom_compx}$a$). On the other hand, for $\alpha=5\times10^{-4}$~bar$^{-0.5}$ and $10\times10^{-4}$~bar$^{-0.5}$ along with $\beta=0.5$, the fraction of SiH$_4$ is mostly larger than our nominal calculation, as expected. In particular, it mostly exceeds 1~\%, and the parameter space of $T_\mathrm{gr}$ and $P_\mathrm{gr}$ with $x_\mathrm{SiH_4}'\geq10$~\% becomes larger than those in our nominal calculations by up to 1000~K in $T_\mathrm{gr}$ and by up to about two orders of magnitude in pressure. Even if the actual water solubility follows that estimated from $\alpha=5\times10^{-4}$~bar$^{-0.5}$ and $10\times10^{-4}$~bar$^{-0.5}$ along with $\beta=0.5$, our model presents a lower
estimate of SiH$_4$ abundance at 0.1~bar.
}

{
In contrast, the boundary of SiH$_4$ presence and absence at 0.1~bar differ from our nominal calculation,  only slightly, as shown in  Fig.~\ref{fig:atom_comp_sol}. 
Compared to the boundary in our nominal calculation (gray lines shown in Fig.~\ref{fig:atom_comp_sol}), the difference is up to a factor of 3 in $P_\mathrm{gr}$ at $T_\mathrm{gr}\sim2500$~K.
Note that the largest difference  occurs under  conditions with $P_\mathrm{H_2, gr}=100$~bar, which is shown by the lowest values in $P_\mathrm{gr}$ not aligning with a Si/O ratio of 1 (see Fig.~\ref{fig:atom_comp_sol}$b$ for $T_\mathrm{gr}\lesssim3100$~K and Fig.~\ref{fig:atom_comp_sol}$c$ for $T_\mathrm{gr}\lesssim3500$~K). It comes from a lower value of $P_\mathrm{gr}$ for a given $P_\mathrm{H_2, gr}$ due to the higher estimate of water solubility than that from our nominal calculation. 
Hence, from the sensitivity tests, we find that the SiH$_4$-rich atmosphere can exist across a broad parameter space in $P_\mathrm{H_2,gr}$ and $T_\mathrm{H_2,gr}$ even adopting the water solubility parameter values derived by other studies \citep{Willson+1981,Berndt+2002,Newcombe+2017, Sossi+2023},
although the SiH$_4$ fraction is sensitive to the adopted parameter values. 
}

\subsection{Condensation of silicates}
\label{sec:d_c}
The condensation of silicates plays a crucial role in removing Si and O from the atmosphere. In our equilibrium calculations, we consider two condensates: SiO and SiO$_2$, which remove Si and O with Si/O ratios of 1 and 0.5, respectively. Our calculations suggest that the condensation of SiO is dominant over that of SiO$_2$, as shown in Fig. \ref{fig:atom_comp1}. Additionally, the pressure boundaries of SiH$_4$ presence at 0.1 bar, as seen in Fig. \ref{fig:atom_compx}$a$, are caused by SiO condensation, which aligns with a Si/O ratio of 1 at the surface.

However, we cannot definitively conclude that SiO condensation is the most dominant process based solely on equilibrium chemistry, as these calculations do not account for the efficiency of condensation and particle growth, which vary depending on the properties of the condensates.  The actual dominant condensate depends on the production, growth, and settling processes of particles, which are determined by the physical properties of the condensates. Exploring these detail is beyond the scope of this work. A recent experiment reported a low sticking coefficient of 0.016 for SiO grains, indicating inefficient growth of SiO in circumstellar outflows \citep{Kimura+2022}.
Even if such inefficient SiO condensation occurs and SiO$_2$ becomes the dominant condensates in the atmospheres we consider, our model presents a lower estimate of SiH$_4$ abundance at 0.1 bar. 

\subsection{Caveats}
\label{sec:d_cm}
\subsubsection{Non-ideality of atmospheric molecules}
\label{sec:d_non}
\begin{figure}[t]
 \begin{minipage}{0.5\textwidth}
    \begin{center}
\includegraphics[width=\textwidth]{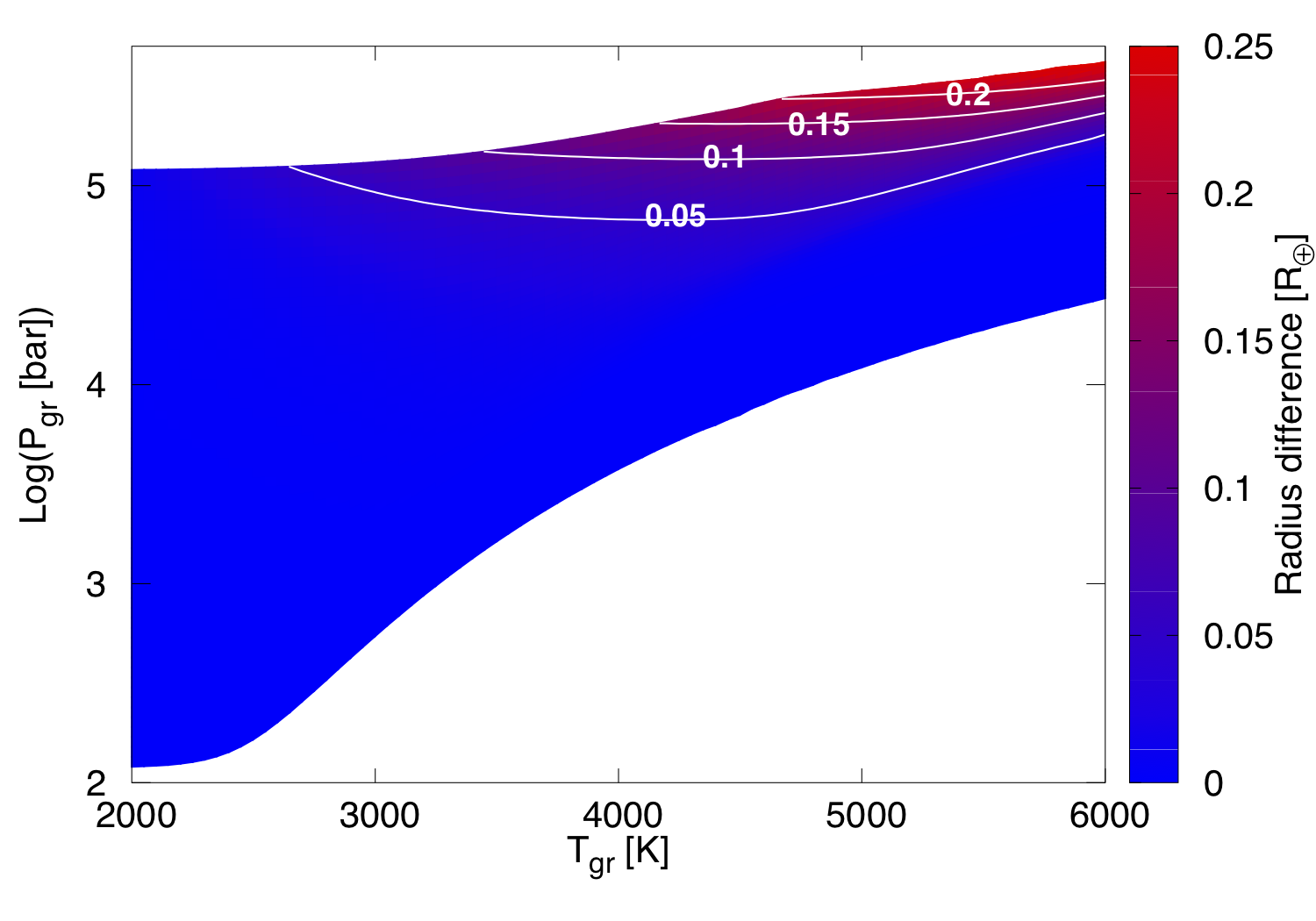}
   \end{center}
 \end{minipage}
\caption{Difference of planetary radius at 0.1~bar between our nominal calculation (Fig.~\ref{fig:atom_compy}) and our calculation adapting non-ideal EOS of CH$_4$ \citep{Zhang+2009} as a proxy for SiH$_4$, for $P_\mathrm{rcb}=10$~bar, under different ground pressure, $P_\mathrm{gr}$, and temperature, $T_\mathrm{gr}$.
}
\label{fig:atom_compr}
\end{figure}

We have not considered the non-ideal behavior of SiH$_4$ in this study because there are no available pubic data on it to our knowledge.
If the non-ideal behavior of SiH$_4$ under high pressure was similar to that of CH$_4$, which is the carbon analogue of SiH$_4$, it could result in a larger planetary radius than our estimates for atmospheres rich in SiH$_4$.
For example, at pressures of $10^4$, and $10^5$~bar, and a temperature of 3000~K, the molar volumes of CH$_4$ exceeds that of an ideal gas by approximately 2 and 10 times, respectively, based on the EOS of CH$_4$ \citep{Zhang+2009}. Such deviations in the molar volumes from ideal gas behavior could also lead to further variations in fugacity coefficient.
Figure~\ref{fig:atom_compr} demonstrates the difference in a planetary radius between our nominal calculation shown in Fig.~\ref{fig:atom_compy} and our calculation with applying non-ideal EOS of CH$_4$ \citep{Zhang+2009} as a proxy for SiH$_4$, for $P_\mathrm{rcb}=10$~bar. Due to the enhanced molar volume, the planetary radius increases by at most 0.2 $R_\oplus$ under the highest pressure condition in our model. However, as molecular properties such as bond length and polarizability differ between SiH$_4$ and CH$_4$ \citep[see][]{CRC2014}, qualitative investigations such as high-pressure experiments and molecular dynamics simulations of the EOS for SiH$_4$ are required for accurately modeling SiH$_4$-rich atmospheres.

\subsubsection{Temperature profiles}
\label{sec:d_tp}
We adopt simplified atmospheric temperature profiles, with isothermal stratospheres above $P_\mathrm{rcb}$ and tropospheres below, parameterized by $P_\mathrm{rcb}$. In this approach, we do not account for the effects of latent heat on the adiabatic lapse rate and thermal conduction, the latter of which may dominate in the deep atmospheric region, as suggested by \citet{Misener+2023}. 
This could be improved by using atmospheric models that incorporate radiative-convective-conductive equilibrium, as formulated in \citet{Misener+2023}. 
If we included these effects in our model, either could enhance the temperature of the deep atmospheric region rich in SiO \citep{Misener+2023}. It would result in higher $T_\mathrm{rcb}$ than our calculations, especially for the atmospheres rich in SiO at their grounds (see Fig.~\ref{fig:atom_compx}$d$).

However, a limitation in modeling the  equilibrium temperature profile, especially in the radiative layer and the conductive layer, arises from the lack of absorption cross-section data for SiH$_4$ at wavelengths shorter than 2 microns \citep{Owen+2017}.
The temperatures of the stratosphere and $P_\mathrm{rcb}$
  generally depend on both the visible and infrared radiative properties of gas species. Additionally, whether thermal conduction is dominant in the deep atmospheric region also depends on opacity \citep{Misener+2023}.
  Therefore, it is currently challenging to determine the radiative-convective-conductive equilibrium for SiH$_4$-rich atmospheres with only its infrared opacity. The SiH$_4$ cross-section decreases with higher transition frequencies in the infrared, suggesting a small optical cross-section for rotational and vibrational transitions, though its electronic transitions may contribute to optical absorption (Yurchenko \& Owens, in private communication). Further research on SiH$_4$ opacity is needed to develop  radiative-convective-conductive equilibrium temperature profiles for SiH$_4$-rich atmospheres.

\subsubsection{Other components possibly affecting SiH$_4$ abundance}
\label{sec:d_oc}
We have shown that the condensation of silicon-oxides through the reaction between SiH$_4$ and H$_2$O is the dominant process for removing Si and reducing the abundance of SiH$_4$ in the atmospheres (Sec.~\ref{sec:res}). While our model focuses on Si-O-H chemistry, specifically the reactions between hydrogen and vaporized gases from SiO$_2$ (R1--R5), we also discuss the potential influence of other atmospheric gas components on SiH$_4$ abundance below.

In sub-Neptune atmospheres, H$_2$O, along with carbon- and nitrogen-bearing gases such as CH$_4$, CO, CO$_2$, NH$_3$, and N$_2$, are expected to be abundant, possibly originating from both nebula gas accretion and outgassing from rocky interiors \citep[e.g.,][]{Miller-Ricci+2012, Shorttle+2024}. SiH$_4$ can react with these species to form silicon carbide (SiC) and silicon nitride (Si$_3$N$_4$) in addition to SiO and SiO$_2$. These compounds, with melting points of 3103~K for SiC and 2173~K for Si$_3$N$_4$, are more refractory than SiO$_2$, which melts at 1995~K \citep{CRC2014}. Consequently, carbon and nitrogen, alongside oxygen, could further reduce SiH$_4$ concentrations through condensation in the atmospheres. Also, SiH$_4$ can react with sulfur-bearing species such as H$_2$S and SO$_2$, which may also be present in sub-Neptunes' atmospheres, producing SiS, a less refractory compound with a melting point of 1363~K \citep{CRC2014}.

As shown in Fig.~\ref{fig:atom_compx}$a$, the molar fraction of SiH$_4$ exceeds 0.1~\% across most of the parameter space we explored. If one assumes that exotic oxygen, carbon, nitrogen and sulfur fractions originating from nebula gas accretion are similar to solar metallicity values \citep[with elemental fractions of O, C, N, and S approximately $5\times10^{-2}$~\%, $3\times10^{-2}$~\%,  $7\times10^{-3}$~\%, and $1\times10^{-3}$~\%, respectively;][]{Lodders2021} and that all of these elements contribute to SiH$_4$ removal via condensation, the atmospheric SiH$_4$ abundance in our simulations would be only slightly altered. In such cases, the SiH$_4$-rich atmospheres would likely lack not only oxygen- but carbon- and nitrogen-bearing gases in their upper layers due to SiC and Si$_3$N$_4$ condensation. Sulfur-bearing gases could also be depleted if the temperature of the upper atmosphere is low enough for SiS condensation.

Outgassing from magma ocean can also supply exotic O, C, N, and S to the atmosphere. For a reduced magma ocean, which is necessary to sustain abundant SiH$_4$ throughout the atmosphere (see Sec.~\ref{sec:d_r}), not only O but also N and S are less likely to be outgassed due to their high solubility into reduced silicate melt  \citep[e.g.,][]{O'Neill+2002, Dasgupta+2022}. However, carbon is more likely to be outgassed than these other volatiles.  For example, on a magma ocean with bulk silicate Earth composition containing H, C, N, and S at $\Delta\mathrm{IW}<-3$, CO and H$_2$ are predicted to be the dominant outgassed species in equilibrium partitioning of C-H-O-N-S between the magma ocean and the atmosphere \citep{Gaillard+2022}.
For the atmospheres of sub-Neptunes with rocky cores containing carbon, \citet{Tian+2024} showed that the molar fraction of CH$_4$ exceeds 10~\% over high-carbon-content magma with a carbon activity, $a_C$, of 0.1 but is only 10$^{-5}$~\% over low-carbon-content magma with $a_C$ of $10^{-7}$ for $\Delta\mathrm{IW}=-3$. Thus, if the magma ocean is reduced yet carbon-rich, SiH$_4$ would react with the abundant CH$_4$, likely forming condensates of SiC and becoming depleted in the atmosphere.

The actual fractions of O, C, N, and S in the rock and accreted gas of sub-Neptunes depend on the specific materials that protoplanets gathered within the protoplanetary disk. The composition of gas and solids in the protoplanetary disk varies in O, C, N, and S content across time and location \citep[see,][for a recent review]{Oberg+2023}. Thus, a more detailed examination requires planetary formation studies investigating the fractions of O, C, N, and S, which will be the focus of a future study.

\subsection{Silane world}
In sub-Neptune atmospheres, refractory elements originating from rocky interiors can react with hydrogen, leading to the formation of various hydrides \citep[][]{Charnoz+2023,Misener+2023,Falco+2024}. Therefore, these metal hydrides could serve as compelling evidence for a highly reduced rocky core within sub-Neptune. Silicon, one of a major rock-forming element, predominantly forms SiH$_4$ as demonstrated in this study. In this section, we discuss that SiH$_4$ can undergo further chemical transformations and that it would coexist with other metal hydrides. 
\subsubsection{Silane family}
\label{sec:d_sf}
\begin{figure}[t]
 \begin{minipage}{0.5\textwidth}
    \begin{center}
\includegraphics[width=\textwidth]{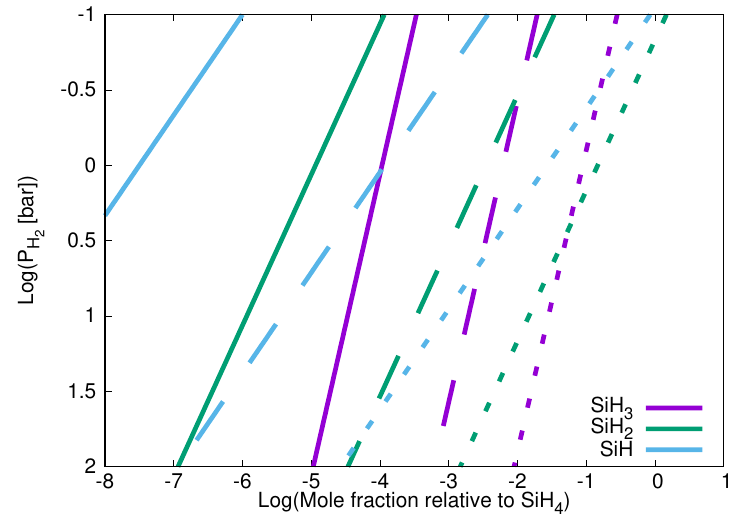}
   \end{center}
 \end{minipage}
\caption{Relative mole fractions of SiH$_3$/SiH$_4$ (purple), SiH$_2$/SiH$_4$ (green), and SiH/SiH$_4$ (light blue) are shown as a function of hydrogen pressure for three isothermal temperatures of 1000~K (solid), 1250~K (dashed) and 
1500~K (dotted).}
\label{fig:sifam}
\end{figure}
Monosilane, SiH$_4$ is one of silicon-hydrides, Si$_a$H$_b$, called silanes. Under an hydrogen-dominated environment, SiH$_4$ would coexist with other silanes.
In the upper layer of the  atmosphere, SiH$_4$ would produce the smaller silanes such as  silanide (SiH$_3$), silylene (SiH$_2$) and silylidyne (SiH), namely,
\begin{equation}
\mathrm{SiH}_{4}  \rightleftharpoons \mathrm{SiH_3} + \frac{1}{2} \mathrm{H_2}. \tag{R6} 
\label{eq:r6}
\end{equation}
\begin{equation}
\mathrm{SiH}_{4}  \rightleftharpoons \mathrm{SiH_2} +  \mathrm{H_2}, \tag{R7} 
\label{eq:r7}
\end{equation}
\begin{equation}
\mathrm{SiH}_{4}  \rightleftharpoons \mathrm{SiH} + \frac{3}{2} \mathrm{H_2}, \tag{R8}  
\label{eq:r8}
\end{equation}

Figure~\ref{fig:sifam} shows their relative fraction to SiH$_4$ in chemical equilibrium for three different temperature of 1000~K (solid), 1250~K (dashed), and 1500~K (dotted) in the pressure range from 100~bar to 0.1~ bar.  These are calculated using the equilibrium constants derived from the thermodynamic data provided in \citet{CHATELAIN+2020}. As shown in Fig.~\ref{fig:sifam}, 
their fractions  increaseas pressure decreases and temperature rises. Specifically, the SiH$_2$/SiH$_4$ ratio exceeds 100~\% at 0.1~bar and 1500~K, compared to about 0.1~\% at 100~bar and 1500K. The pressure dependencies are written as follows; SiH$_3$/SiH$_4$$\propto P_\mathrm{H_2}^{-0.5}$, SiH$_2$/SiH$_4$$\propto P_\mathrm{H_2}^{-1}$ and SiH/SiH$_4$$\propto P_\mathrm{H_2}^{-1.5}$.
Thus, the smaller silanes could dominate over SiH$_4$ at the upper layer of the atmosphere with $T_\mathrm{rcb}\geq1500$~K, corresponding to our calculations with high $T_\mathrm{gr}$ and high $P_\mathrm{gr}$ for $P_\mathrm{rcb}=10$ and 100~bar (see Fig.~\ref{fig:atom_compy}$a$ and $b$).

In addition, under SiH$_4$-rich conditions, SiH$_4$ could produce the larger silanes such as disilane (Si$_2$H$_6$) and trisilanes (Si$_3$H$_8$), namely,
\begin{equation}
2\mathrm{SiH}_{4}  \rightleftharpoons \mathrm{Si_2H_6} + \mathrm{H_2}, \tag{R9}  
\label{eq:r9}
\end{equation}
\begin{equation}
3\mathrm{SiH}_{4}  \rightleftharpoons \mathrm{Si_3H_8} +  2\mathrm{H_2}. \tag{R10}  
\label{eq:r10}
\end{equation}
We also estimate their relative fractions to SiH$_4$ in chemical equilibrium
based on the thermodynamic data in \citet{CHATELAIN+2020}.
Under conditions with the SiH$_4$ molar fraction of 1~\% and temperature of 500~K, 750~K and 1000~K,
 the relative fraction of Si$_2$H$_6$ is about 0.3, 1 and 3~\%, respectively, while that of Si$_3$H$_8$ is about $4\times10^{-4}$, $5\times10^{-3}$ and 0.02~\%, respectively.
The equilibrium fractions of Si$_2$H$_6$ and Si$_3$H$_8$ relative to SiH$_4$ increase proportionally to $x_\mathrm{SiH_4} x_\mathrm{H_2}^{-1}$ and $x_\mathrm{SiH_4}^2x_\mathrm{H_2}^{-2}$, respectively, but do not depend on pressure, which differs from the behaviors of SiH$_3$, SiH$_2$ and SiH shown in Fig~\ref{fig:sifam}. Thus, the atmospheres with the SiH$_4$ fraction of $\geq10$~\% likely contain the large amount of Si$_2$H$_6$ and Si$_3$H$_8$ as well. 

In this section above, we have discussed the abundances of silanes in chemical equilibrium, which is generally archived in the high pressure regions of atmospheres. However, reactions between SiH$_4$ and radicals such as the smaller silanes, atomic hydrogen and ions produced by photochemistry could lead to further variety of silanes including cyclic compounds such as  cyclotetrasilane (Si$_4$H$_8$), cyclopentasilane (Si$_5$H$_{10}$) and cyclohexasilane (Si$_6$H$_{12}$) \citep[e.g.,][]{{DeBleecker+2004}}.
Such higher-order silane compounds and silane particles are known to be also produced via the decomposition of SiH$_4$ in hydrogen gases \citep[][and references therein]{Onischuk+2001}. 
Monosilane pyrolysis experiments indicate that decomposition is minimal below ${450}^\circ$C (723K) but starts at higher temperatures, producing higher-order silanes at 1 atm \citep{WYLLER2017, WYLLER+2020}.  This decomposition and the subsequent nucleation of silane particles are expected to occur if the SiH$_4$ concentration exceeds a critical threshold. The critical concentration of SiH$_4$ decreases with temperature: approximately 10~\%, 1~\%, and 0.2~\% at 700~K, 900~K, and 1400~K, respectively, at 1 atm \citep{WYLLER2016, MURTHY19761}.
A detailed investigation of silanes other than SiH$_4$ is beyond the scope of this study and will be addressed in  future research.

\subsubsection{Other metal hydrides}
In addition to silanes, other metal hydrides may form through the vaporization of rocky cores and the reactions between rocky vapors and hydrogen in sub-Neptune atmospheres. Major rock-forming elements such as Na, K, Fe, and Mg are predicted to form NaH, KH, FeH, and MgH in the deeper atmospheric layers of giant planets, brown dwarfs, and sub-Neptunes, according to chemical equilibrium calculations \citep[e.g.,][]{Fegley+1994, Visscher+2010, Charnoz+2023, Falco+2024}. Laser-heated diamond-anvil cell experiments in the MgO-Fe-H$_2$ system also indicate the formation of FeH$_a$ alloy at 27--40 GPa and 2700--3100 K \citep{Horn+2023}, and the formation of Mg$_2$FeH$_6$, Mg(OH)$_2$, FeH$_a$ alloy, and H$_2$O at 8--13 GPa and temperatures above 3500 K, which is close to or exceeding the melting temperature of MgO \citep{Kim+2023}.  
 Therefore, SiH$_4$ would coexist with these hydrides in sub-Neptunes with rocky cores.
 Future observations of not only SiH$_4$ but also the hydrides of other rocky elements in sub-Neptune atmospheres could help constrain the element abundances of their  magma oceans. 

\subsubsection{Sub-Neptune's property with SiH$_4$-rich atmospheres}
\label{sec:sns}
 In the parameter space we explored over $P_\mathrm{H_2,gr}$ and $T_\mathrm{gr}$, the planetary radii and temperatures at 0.1~bar of sub-Neptunes with SiH$_4$-rich atmospheres distribute in
 $1.7 \leq  R_p/R_{\oplus} \leq 2.4$ 
 and $T_\mathrm{rcb}\leq$2000~K 
 for $P_\mathrm{rcb}=100$~bar, 
 $1.6 \leq  R_p/R_{\oplus} \leq 2.2$ 
 and $T_\mathrm{rcb}\leq$1500~K 
 for $P_\mathrm{rcb}=10$~bar,
  $1.6 \leq  R_p/R_{\oplus} \leq 2.2$ 
 and $T_\mathrm{rcb}\leq$1000~K 
 for $P_\mathrm{rcb}=1$~bar,
 as shown in Fig.~\ref{fig:atom_compy}.
 Note that our calculations might provide a lower estimate in the planetary radii for the SiH$_4$-rich atmospheres, as discussed in Sec.~\ref{sec:d_non}. 
 
 The atmospheres with $T_\mathrm{rcb}$ below 723~K would have the abundant 
 SiH$_4$ shown in our results (see Fig.~\ref{fig:atom_compx}$a$), while 
 the higher temperature atmosphere may have various silanes formed through the decomposition of SiH$_4$ rather than only SiH$_4$, as discussed in Sec.~\ref{sec:d_sf}.
The relation between an irradiation field and temperature of the upper layers of SiH$_4$-rich atmosphere remains unclear due to the lack of opacity data of SiH$_4$ in optical but it will be addressed in future research, as discussed in Sec.~\ref{sec:d_tp}. 

\subsection{Observational implications}
\begin{figure}[t]
 \begin{minipage}{0.5\textwidth}
    \begin{center}
\includegraphics[width=\textwidth]{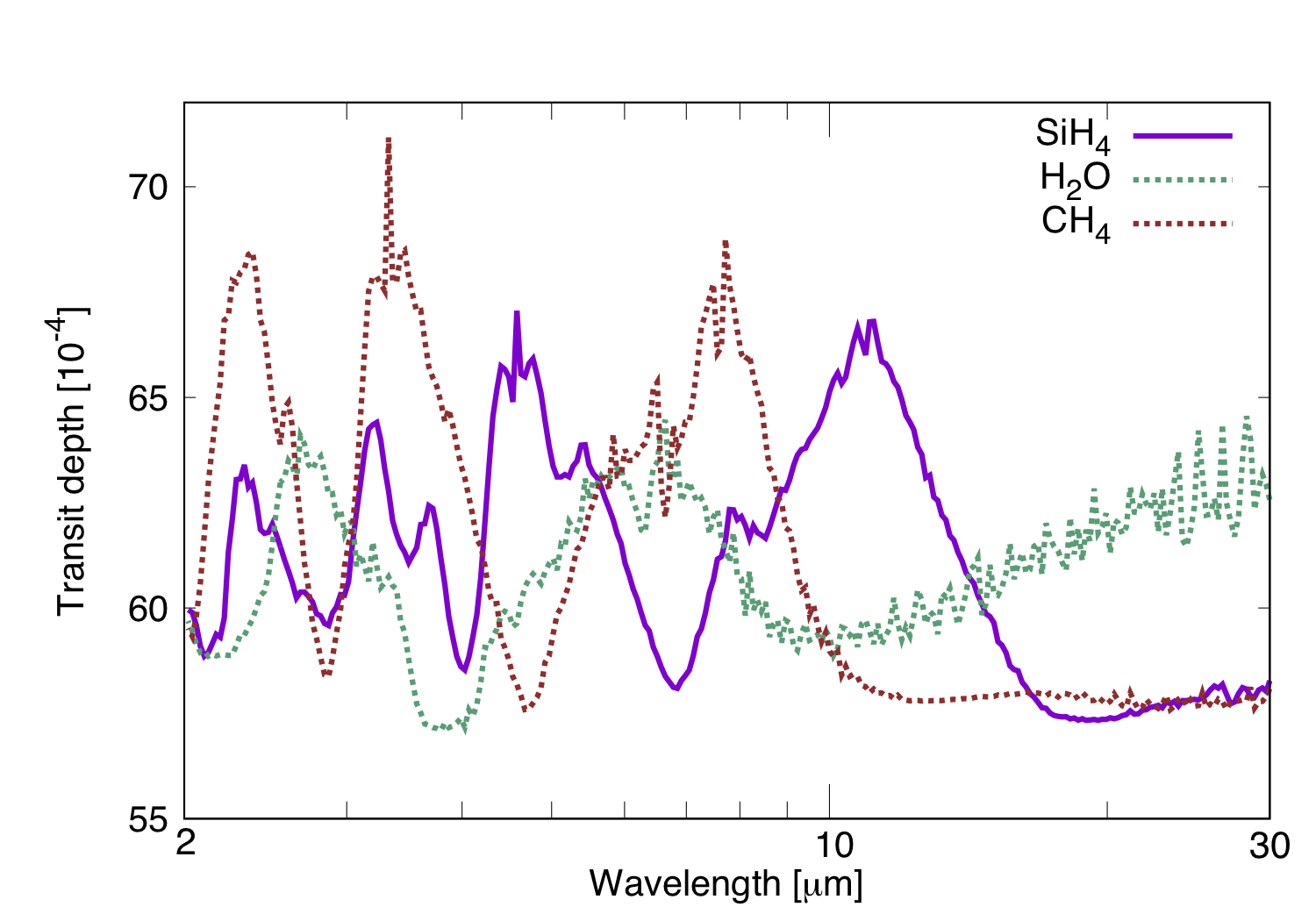}
   \end{center}
 \end{minipage}
\caption{Transmission spectrum for the atmosphere of a planet orbiting a star with a radius of 0.2 $R_\odot$ under the same conditions as in Fig.~\ref{fig:atom_comp1}
, with a SiH$_4$ fraction of 7~\%, a tropopause pressure of 10bar, a surface H$_2$ pressure of 3$\times10^4$bar, and a surface temperature of 3000K (solid purple line).
  The dotted lines represent the transmission spectra for atmospheres rich in H$_2$O (green) or CH$_4$ (brown) under the same condition with the SiH$_4$-rich atmosphere, adding offset ($-5\times10^{-4}$ and $-7\times10^{-4}$, respectively) for comparison. 
  }
\label{fig:transm}
\end{figure}

The SiH$_4$-rich atmospheres predicted in this study are expected to show the features of SiH$_4$ in spectroscopic measurements, which were not indicated in  \citet{Falco+2024} due to the absence of H$_2$O dissolution effects in their atmospheric model.
Figure.~\ref{fig:transm} illustrates the transmission spectrum of a SiH$_4$-rich atmosphere for a planet orbiting a star with a radius of 0.2 $R_\odot$, under the same conditions as in Fig.~\ref{fig:atom_comp1}. To compute the transmission spectrum, we used the open-source code \textit{TauREx 3} \citep{Al-Refaie+2021}. 

The spectrum exhibits four unique SiH$_4$ signal bands with wavelength variations of several hundreds ppm in the ranges 2--3~$\mu$m, 3--4~$\mu$m, 4--7~$\mu$m, and 7--18~$\mu$m. The spectral feature of the SiH$_4$-rich atmosphere is distinguishable from those of an H$_2$O-rich atmosphere (green dotted) and an CH$_4$-rich atmosphere (brown dotted), as shown in Fig.~\ref{fig:transm}.
These unique SiH$_4$ features provide a promising target for future observations, particularly with infrared instruments capable of detecting molecular absorption bands. Ongoing JWST and upcoming missions such as Ariel \citep{Tinetti+2022} with their enhanced sensitivity and spectral resolution are  well-suited for identifying SiH$_4$ in sub-Neptunes' atmospheres. Detecting these spectral features would not only confirm the presence of SiH$_4$ but also offer new insights into the atmospheric chemistry of sub-Neptunes, revealing the evidence of interactions between their rocky cores and hydrogen-dominated atmospheres.

\subsection{Relation to previous observations of sub-Neptune atmospheres}
Previous works observed transmission spectra for various sub-Neptunes.
Those observations suffered from high-altitude clouds and hazes \citep[e.g.,][]{Kreidberg+14,Knutson+14,Brande+24}. 
JWST has begun to shed light on the nature of sub-Neptune atmospheres, including the presence of multiple molecules such as H$_2$O, CH$_4$, CO$_2$, and SO$_2$ \citep[][]{Madhusudhan+2023,Benneke+2024,Holmberg+2024,Beatty+24,Davenport+25} and the possible prevalence of highly metal-rich atmospheres \citep{Kempton+23,Gao+23,Schlawin+24,Piaulet+24,Ohno+25}, although SiH$_4$ detection has not been reported yet.

Most of the currently observed sub-Neptunes unlikely belong to the novel silane world regime proposed in this study, since they exhibit signature of H$_2$O and/or other carbon-bearing molecules that do not coexist with abundant SiH$_4$ according to our result.
Nondetection of the silane world may indicate that many sub-Neptunes have FeO-rich rocky cores; otherwise, they are icy planets migrated from outer orbits.
On the other hand, several sub-Neptunes and super-Earths show featureless spectra that are attributed to high mean molecular weight and/or high-altitude aerosols \citep[e.g.,][]{Alderson+24,Wallack+24,Scarsdale+24,Alam+25}, and it would be worthwhile to examine the possibility of the silane world.
Although the universality/rarity of the silane world remains an open question, if detected, they will provide a compelling evidence for the highly reduced rocky core.

\section{Summary and Conclusion}
\label{sec:sum}
We investigate the atmospheric composition of sub-Neptunes with  reduced FeO-free magma oceans using an one-dimensional atmospheric model based on the chemical equilibrium of H-/O-/Si-bearing species. This model incorporates the vaporization of SiO$_2$, silicate condensation, and the dissolution of H$_2$O into the magma ocean. SiH$_4$ would likely be formed in the atmosphere overlying a reduced magma ocean, as suggested in chemical equilibrium calculations \citep{Charnoz+2023,Misener+2023} and experimental studies \citep{Shinozaki+2014,Shinozaki+2016}. We find that the dissolution of H$_2$O into the magma ocean \citep{Kennedy1962, Holtz+2000} can  further reduce the atmospheres, allowing abundant SiH$_4$ to persist throughout the atmosphere by preventing its reversion to silicates (Section~\ref{sec:res}). 

Our results suggest that the presence of SiH$_4$ in the upper atmosphere of sub-Neptunes indicates interactions between the atmospheres and a highly reduced magma ocean. Although the existence and redox state of magma oceans in sub-Neptunes remain unknown, detecting SiH$_4$ in future observations could provide insights into their building blocks, core differentiation, and Si speciation in gas-melt-core systems.
Future laboratory and numerical 
studies on SiH$_4$'s properties, such as non-ideal behavior and opacity in the visible and near-infrared, will be quite
helpful for improving the atmospheric models of hydrogen-dominated atmospheres interacting with reduced magma oceans in sub-Neptunes.

\begin{acknowledgments}
We appreciate Sergey Yurchenko and Alec Owens for giving fruitful discussion about the opacity of SiH$_4$. We also appreciate the referee, Kevin Heng, for his careful reading and valuable comments which helped to improve this paper greatly. We also thank Giovanna Tinetti for helpful discussions.
This work was supported by JSPS KAKENHI grant No. 22K14090. YI and YF were also supported by JSPS KAKENHI grant No. 25K01062.
\end{acknowledgments}
\bibliography{ref}{}
\bibliographystyle{aasjournal}

\appendix
\section{Parameter space where water fraction in magma reaches its upper limit}
\begin{figure}[t]
    \begin{center}
 \begin{minipage}{0.5\textwidth}
\includegraphics[width=\textwidth]{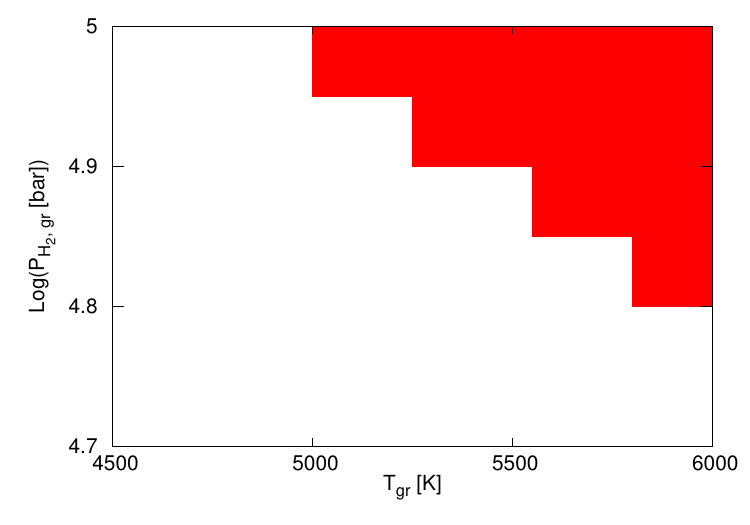}
 \end{minipage}
    \end{center}
\caption{The parameter space of ground temperature and hydrogen ground pressure where $X_\mathrm{H_2O}$
 reaches the upper limit in our results is shown as a red shaded region.}
\label{fig:limit}
\end{figure}

We set the upper limit of a H$_2$O mass fraction in magma, $X_\mathrm{H_2O}$, at 10~wt\% in our model to avoid the artificial extrapolation of H$_2$O solubility law (Eq. \ref{eq:diss}) to conditions where magma oceans become fully miscible. Figure~\ref{fig:limit} shows the parameter space where $X_\mathrm{H_2O}$ reaches its upper limit. $X_{\mathrm{H_2O}}$ attains this upper limit only within a narrow range of high surface temperatures ($T_{\mathrm{gr}} \geq 5000$ K) and hydrogen ground pressures ($P_{\mathrm{H_2,gr}} \geq 10^{4.8}$ bar). It suggests that the magma may become fully miscible with water under these conditions, though a detailed investigation of such states is beyond the scope of this study.

\section{Numerical way to find equilibrium solution at ground}
\begin{figure}[t]
    \begin{center}
 \begin{minipage}{0.5\textwidth}
\includegraphics[width=\textwidth]{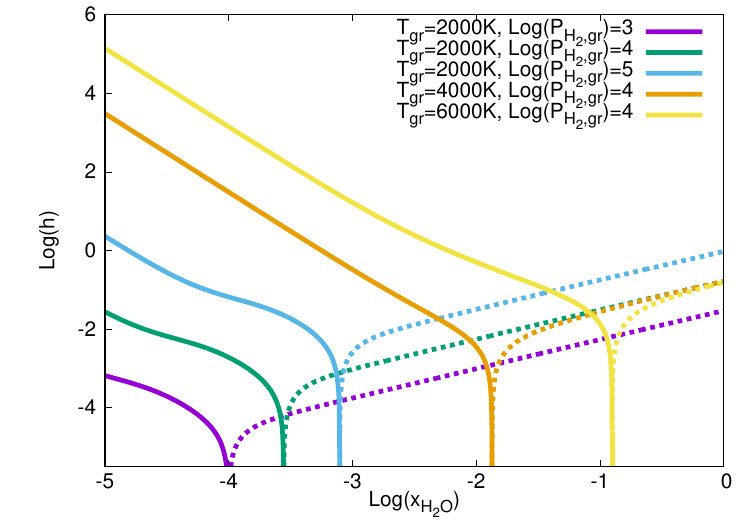}
 \end{minipage}
    \end{center}
\caption{$h$ (Eq.~\ref{eq:a1}) are shown as functions of $x_\mathrm{H_2O}$ for five combinations of ground temperature and hydrogen ground pressure; $T_\mathrm{gr}=2000$~K and $P_\mathrm{H_2,gr}=10^3$~bar (purple), 
2000~K and $10^4$~bar (green),
2000~K and $10^5$~bar (cyan),
4000~K and $10^4$~bar (orange),
and 6000~K and $10^4$~bar (yellow).
Dotted lines represent $-h$, the function $h$ multiplied by $-1$.}
\label{fig:func}
\end{figure}

As described in Sec.\ref{sec:mod}, we derive  equilibrium composition at ground for given $P_\mathrm{H_2,gr}$ and $T_\mathrm{gr}$ by finding the value of $x_\mathrm{H_2O}$ that satisfies Eq.~(\ref{eq:feqg}).
Using Eqs.~(\ref{eq:f})--(\ref{eq:Mh2}), the equation (\ref{eq:feqg}) can be rewritten as,
\begin{eqnarray}
    h(x_\mathrm{H_2O}, P_\mathrm{H_2, gr}, T_\mathrm{gr})&=& \frac{36 \pi R_\mathrm{core}^4}{\zeta GM_\mathrm{core}^2}  (\chi_\mathrm{H_2}^{-1} + \Sigma_i x_i) 
    \left(\frac{1}{1-Y_\mathrm{He}}+\Sigma_i x_i \frac{m_i}{m_\mathrm{H_2}}
    \right)^{-1} \notag \\
    && \times P_\mathrm{H_2, gr} (2x_\mathrm{SiH_4}+x_\mathrm{SiO}-x_\mathrm{H_2O}-2x_\mathrm{O_2}) - \alpha 
{\left( \frac{P_\mathrm{H_2, gr}}{\tilde{P_0}}\right)^\beta}    
x_\mathrm{H_2O}^\beta, \notag \\
    &=&0.
    \label{eq:a1}
\end{eqnarray} 
$h$ is the function of $x_\mathrm{H_2O}$ for given $P_\mathrm{H_2,gr}$ and $T_\mathrm{gr}$ (see also Eqs.~\ref{eq:x_sio}--~\ref{eq:x_sih4}). 
Figure~\ref{fig:func} shows the function, $h$, for five combinations of ground temperature and hydrogen ground pressure; $T_\mathrm{gr}=2000$~K and $P_\mathrm{H_2,gr}=10^3$~bar (purple), 
2000~K and $10^4$~bar (green),
2000~K and $10^5$~bar (cyan),
4000~K and $10^4$~bar (orange),
and 6000~K and $10^4$~bar (yellow). As shown in Fig~\ref{fig:func}, $h$ decreases continuously with $x_\mathrm{H_2O}$ for all the given $T_\mathrm{gr}$ and $P_\mathrm{H_2,gr}$. 
This continuous decrease of $h$ with $x_\mathrm{H_2O}$ is confirmed by parameterizing $x_\mathrm{H_2O}$ from 10$^{-18}$ to 1 for $T_\mathrm{gr}=2000$--6000~K and $P_\mathrm{H_2,gr}=10^2$--$10^5$~bar. Since $h$ is positive at $x_\mathrm{H_2O}=0$ (see Eq.~\ref{eq:a1}), we numerically obtain the solution for $h=0$ by gradually increasing $x_\mathrm{H_2O}$.

\section{Approximated solution of pressure boundaries for SiH$_4$ presence}
We find the pressure boundaries with a sharp decline in the SiH$_4$
abundance at 0.1 bar, as shown in Fig~\ref{fig:atom_compx}. As explained in Sec.\ref{sec:res}, the boundaries are caused by the condensation of SiO(c), aligning with a Si/O ratio of 1 at the
ground.
Although we cannot conclude that the condensation of SiO(c) dominantes over that of SiO$_2$(c) from our calculation, we derive the approximated solution of the pressure boundaries below.

At the pressure boundary 
with a Si/O ratio of one, the  faction of SiH$_4$ relative to H$_2$ at ground approximately equals to that of H$_2$O as the molar fraction of O$_2$ is negligible compared to them. Therefore, in the case, from Eq.~(\ref{eq:x_sih4}), the  fraction of H$_2$O can be written as
\begin{eqnarray}
    x_\mathrm{H_2O}=(K_\mathrm{eq,R1}K_\mathrm{eq,R2}K_\mathrm{eq,R3}P_\mathrm{H_2,gr}{/P_0})^{\frac{1}{3}}.
    \label{eq:a2}
\end{eqnarray}
Additionally, to derive the approximated solution, we assume SiO is the most dominant gas species at ground in the atmospheres. Based on our results, this assumption is reasonable for $T_\mathrm{gr}\geq3000$~K (see Fig.~\ref{fig:atom_compx}$d$ and $f$).
By assuming $x_\mathrm{SiO} \gg x_\mathrm{SiH_4}(=x_\mathrm{H_2O}) \gg x_\mathrm{O_2}$, Eq.~(\ref{eq:a2}), and $P_\mathrm{gr}\sim P_\mathrm{H_2,gr}x_\mathrm{SiO}$ in Eq.~(\ref{eq:a1}),
one can obtain the approximated solution of the pressure boundaries with a Si/O ratio of one as
\begin{eqnarray}
    { \frac{P_\mathrm{H_2, gr}}{{P_0}}}
    &\sim& \left[ \left(
    \frac{18 {P_0} \pi R_\mathrm{core}^4}{11 \alpha {\left( \frac{P_0}{\tilde{P_0}}\right)^\beta} \zeta GM_\mathrm{core}^2} \right )^{3}
    K_\mathrm{eq,R1}^{2-\beta}
    K_\mathrm{eq,R2}^{2-\beta}
    K_\mathrm{eq,R3}^{-1-\beta} 
 \right]^\frac{1}{1+4\beta},
 \label{eq:a3}
 \\
    { \frac{P_\mathrm{gr}}{{P_0}}}
    &\sim&  \left[ \left(
    \frac{18 {P_0}
    \pi R_\mathrm{core}^4}{11 \alpha     {\left( \frac{P_0}{\tilde{P_0}}\right)^\beta}
    \zeta GM_\mathrm{core}^2} \right )^{-1}
    K_\mathrm{eq,R1}^{3\beta}
    K_\mathrm{eq,R2}^{3\beta}
    K_\mathrm{eq,R3}^{-\beta} \right]^\frac{1}{1+4\beta}.
     \label{eq:a4}
\end{eqnarray}

\begin{figure}[t]
    \begin{center}
 \begin{minipage}{0.5\textwidth}
\includegraphics[width=\textwidth]{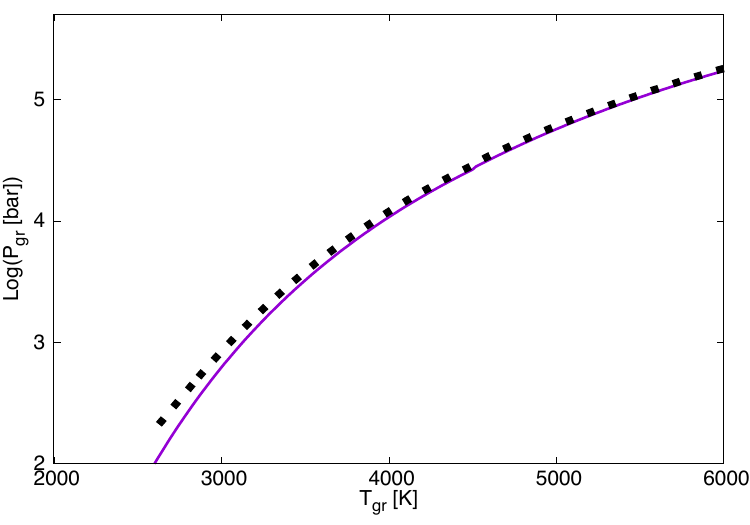}
 \end{minipage}
    \end{center}
\caption{Approximated solution of pressure boundaries with a Si/O ratio of one is shown as a function of ground temperature for parameters assumed in our nominal simulation in Sec.~\ref{sec:res}.
Dotted line represents the pressure boundaries calculated by our model, which is the same as that shown in Fig.~\ref{fig:atom_compx}.}
\label{fig:func2}
\end{figure}

Figure~\ref{fig:func2} shows the approximated solution described by Eq.(\ref{eq:a4}) for the same parameters (i.e., $\alpha, \beta, \zeta, M_\mathrm{core}, R_\mathrm{core}$) assumed in our nominal simulation in Sec.\ref{sec:res}. The approximated solution shows better agreement with the numerical solution (dotted line in Fig.~\ref{fig:func2}) at higher temperatures.

\end{document}